\documentclass[journal]{IEEEtran}
\usepackage{times}
\usepackage{epsfig}
\usepackage{graphicx}
\usepackage{amsmath}
\usepackage{amssymb}
\usepackage{colortbl}%añadidoooooo color tablaaaa
\usepackage{multirow}
\usepackage{soul}
\usepackage{adjustbox}
\usepackage{array}
\usepackage{setspace}
\usepackage{lscape}
\usepackage{mathtools}
\usepackage{hyperref}

\usepackage{pgfplots}
\pgfplotsset{compat=newest}

\usepackage{enumitem}
\usepackage{array}
\usepackage{multirow}
\usepackage{rotating}
\usepackage{caption}
\usepackage{cite}
\usepackage{amsmath,amssymb,amsfonts}
\usepackage{algorithmic}
\usepackage{graphicx}
\usepackage{textcomp}
\usepackage{subfigure}
\usepackage{caption} % DO NOT CHANGE THIS AND DO NOT ADD ANY OPTIONS TO IT

\ifCLASSINFOpdf
\else
\fi
\begin{document}
%\title{BioTouchPass2: Biometric Passwords using Time-Aligned Touch Interaction and Recurrent Neural Networks}
\title{Child-Computer Interaction with Mobile Devices: Recent Works, New Dataset, and Age Detection}

%: Understanding Child-Computer Interaction


\author{Ruben Tolosana, Juan Carlos Ruiz-Garcia, Ruben Vera-Rodriguez, Jaime Herreros-Rodriguez,\\ Sergio Romero-Tapiador, Aythami Morales, Julian Fierrez\\
        % <-this % stops a space

%\author{Michael~Shell,~\IEEEmembership{Member,~IEEE,}
%        John~Doe,~\IEEEmembership{Fellow,~OSA,}
%        and~Jane~Doe,~\IEEEmembership{Life~Fellow,~IEEE}% <-this % stops a space

\thanks{R. Tolosana, J.C. Ruiz-Garcia, R. Vera-Rodriguez, S. Romero-Tapiador, A. Morales and J. Fierrez are with the Biometrics and Data Pattern Analytics - BiDA Lab, Escuela Politecnica Superior, Universidad Autonoma de Madrid, 28049 Madrid, Spain (e-mail: ruben.tolosana@uam.es; ruben.vera@uam.es; juanc.ruiz@uam.es; sergio.romero@uam.es; aythami.morales@uam.es; julian.fierrez@uam.es).

J. Herreros-Rodriguez is with the Hospital Universitario Infanta Leonor, 28031 Madrid, Spain (e-mail: hrinvest@hotmail.com).}}% <-this % stops a space

\maketitle

\begin{abstract}
This article provides an overview of recent research in Child-Computer Interaction with mobile devices and describe our framework ChildCI intended for: \textit{i)} overcoming the lack of large-scale publicly available databases in the area, \textit{ii}) generating a better understanding of the cognitive and neuromotor development of children along time, contrary to most previous studies in the literature focused on a single-session acquisition, and \textit{iii)} enabling new applications in e-Learning and e-Health through the acquisition of additional information such as the school grades and children's disorders, among others. Our framework includes a new mobile application, specific data acquisition protocols, and a first release of the ChildCI dataset\footnote{\url{https://github.com/BiDAlab/ChildCIdb_v1}} (ChildCIdb v1), which is planned to be extended yearly to enable longitudinal studies. 

In our framework children interact with a tablet device, using both a pen stylus and the finger, performing different tasks that require different levels of neuromotor and cognitive skills. ChildCIdb is the first database in the literature that comprises more than 400 children from 18 months to 8 years old, considering therefore the first three development stages of the Piaget’s theory. In addition, and as a demonstration of the potential of the ChildCI framework, we include experimental results for one of the many applications enabled by ChildCIdb: children age detection based on device interaction. 
\end{abstract}


% \fi
\IEEEpeerreviewmaketitle

\section{Introduction}
\IEEEPARstart{C}{hildren} are becoming one of the latest (and youngest) users of the technology based on touch interaction. They have more and more access to mobile devices on a daily basis. This fact is demonstrated in recent studies of the literature~\cite{kilicc2019exposure}, showing that over 75.6\% of the children are exposed to mobile devices between the age of 1 to 60 months. This aspect has been exacerbated by the COVID-19 outbreak in 2020. With a large percentage of the academic institutions around the world now in lockdown, virtual education has temporally replaced traditional education to a very large extent using specific e-Learning mobile applications in which children interact with them to improve their knowledge and skills~\cite{kucirkova2020lessons}. However, and despite the importance of the topic, the field of Child-Computer Interaction (CCI) is still in its infancy~\cite{tsvyatkova2019review}.

Our work aims at generating a better understanding of the way children interact with mobile devices during their development process. Children undergo many different physiological and cognitive changes as they grow up, which reflect in the way they understand and interact with the environment. According to Piaget's theory~\cite{piaget2008psychology}, there are four different stages in the development of the children: \textit{i) Sensorimotor} (from birth to 2 years old), focused mainly on the evolution of the motor control such as fingers and gestures, and the acquisition of knowledge through sensory experiences and manipulating objects; \textit{ii) Preoperational} (2-7 years), children are getting better with language and thinking, improving also their motor skills; \textit{iii) Concrete Operational} (7-11 years), their thinking becomes more logical and organized, but still very concrete; and \textit{iv) Formal Operational} (adolescence to adulthood), they begin to think more about moral, philosophical, ethical, social, and political issues that require theoretical and abstract reasoning. 

Currently, most studies in the field of CCI are focused on the Preoperational and Concrete Operational stages (2-11 years), pointing out that children's touch interaction patterns are different compared with adults~\cite{vatavu2015child,anthony2016gestures,2019_LognormalityChapter_Vera,2018_IETB_DetectChildTouch_Acien}. As a result, different guidelines should be considered for the proper design and development of children mobile applications, considering their incipient physiological and cognitive abilities~\cite{mcknight2010children,anthony2014designing,vatavu2015touch,lanna2019touch}.

In this article we present our framework named ChildCI, which is mainly intended for: \textit{i)} overcoming the lack of large-scale publicly available databases in the area, \textit{ii}) generating a better understanding of the cognitive and neuromotor development of children along time, contrary to most previous studies in the literature focused on a single-session acquisition, and \textit{iii)} enabling new applications in e-Learning~\cite{2020_COMPSAC_HRedBB_JHO} and e-Health~\cite{2019_FG_PDhandw_Castrillon} through the acquisition of additional information such as the school grades and children's disorders, among others. In particular, the present study introduces all the details regarding the design and development of a new child mobile application, the specific acquisition protocol considered, and the first capturing session of the ChildCI database (ChildCIdb v1). In the scenario considered, children interact with a tablet device, using both a pen stylus and also the finger, performing different tasks that require different levels of neuromotor and cognitive skills. Unlike most previous studies in the literature, our analysis considers the first three stages of the Piaget's theory in order to perform an in-depth analysis of the children development process. Additionally, ChildCI is an on-going project in which children will be captured in multiple sessions along their development process (from 18 months to 8 years old), being possible to extract very relevant insights. 

The main contributions of this study are as follow:

\begin{itemize}
\item An overview of recent works studying touch and stylus interactions performed by children on screens, remarking the publicly available datasets for research in this area and the improvements over them of our contributed ChildCIdb.
\item Design and development of a novel child mobile application composed of 6 tests grouped in two different categories (touch and stylus). Different levels of neuromotor and cognitive skills are required in each test to measure the evolution of the children in each Piaget's stage. By doing so, we are able to study the children's performance on finger and stylus in relation with their level of cognitive and motor development according to their age.   
\item A first release of the new ChildCI dataset\footnote{\url{https://github.com/BiDAlab/ChildCIdb_v1}} (ChildCIdb v1), which is planned to be extended yearly to enable longitudinal studies. This is the largest publicly available dataset to date for research in this area with 438 children in the ages from 18 months to 8 years old. In addition, the following aspects are considered in the acquisition of the dataset: \textit{i)} interaction with screens using both finger and pen stylus, \textit{ii)} information regarding the previous experience of the children with mobile devices, \textit{iii)} the children's grades at the school, and \textit{iv)} information regarding the attention-deficit/hyperactivity disorder (ADHD).
\item Example application using machine learning techniques to demonstrate the research potential of our contributed ChildCIdb. In particular, we focus on the task of children age group detection while colouring a tree (named Drawing Test). A new set of 34 global features are proposed to automatically detect the age group, achieving interesting insights.
\end{itemize}

The remainder of the article is organised as follows. Sec.~\ref{related_works} summarises previous studies carried out in touch and stylus interactions performed by children. Sec.~\ref{sec:ChildD} describes all the details of ChildCIdb, including the design and development of the mobile application, the specific acquisition protocol, and the first capturing session. Sec.~\ref{sec:ChildD_applications} develops an example application using machine learning techniques and ChildCIdb for the task of children age group detection. Finally, Sec.~\ref{conclusions} draws the final conclusions and points out future work.

\begin{table*}[t]
\centering
\caption{Comparison of different studies focused on the interaction of the children with mobile devices.}
\scalebox{0.88}{
\begin{tabular}{ccccc}
\textbf{Study}                                                                      & \textbf{Age of Participants} & \textbf{\# Participants} & \textbf{Acquisition Tool} & \textbf{Public Database [ref]} \\ \hline
\begin{tabular}[c]{@{}c@{}}Crescenzi and Grané (2019)\\ \cite{lanna2019touch}\end{tabular}        & 14-33 Months                 & 21                       & Finger                                 & No                 \\
\begin{tabular}[c]{@{}c@{}}Nacher \textit{et al.} (2015)\\ \cite{nacher2015multi}\end{tabular}              & 2-3 Years                    & 32                       & Finger                                & No                 \\
\begin{tabular}[c]{@{}c@{}}Hiniker \textit{et al.} (2015)\\ \cite{hiniker2015touchscreen} \end{tabular}             & 2-5 Years                    & 34                       & Finger                                 & No                 \\
\begin{tabular}[c]{@{}c@{}}Abdul-Aziz (2013)\\ \cite{aziz2013children}\end{tabular}                 & 2-12 Years                   & 33                       & Finger                                 & No                 \\
\begin{tabular}[c]{@{}c@{}}Vatavu \textit{et al.} (2015)\\ \cite{vatavu2015touch}\end{tabular}         & 3-6 Years                    & 89                       & Finger                                  & Yes\cite{vatavu2015touch}                \\
\begin{tabular}[c]{@{}c@{}}Vera-Rodriguez \textit{et al.} (2020)\\ \cite{2019_LognormalityChapter_Vera}\end{tabular}    & 3-6 Years                    & 89                       & Finger                    & Yes\cite{vatavu2015touch}            \\
\begin{tabular}[c]{@{}c@{}}Acien \textit{et al.} (2019)\\ \cite{2018_IETB_DetectChildTouch_Acien}\end{tabular}               & 3-6 Years                    & 89                       & Finger                                & Yes\cite{vatavu2015touch}                       \\
\begin{tabular}[c]{@{}c@{}}Remi \textit{et al.} (2015)\\ \cite{remi2015exploring}\end{tabular}                & 3-6 Years                    & 60                       & Stylus                     & No                 \\
\begin{tabular}[c]{@{}c@{}}Behnamnia \textit{et al.} (2020)\\ \cite{BEHNAMNIA2020105227}\end{tabular}           & 3-6 Years                    & 7                        & Finger                              & No                 \\
\begin{tabular}[c]{@{}c@{}}Hussain \textit{et al.} (2016)\\ \cite{hussain2016children}\end{tabular}            & 4-6 Years                    & 10                       & Finger                                & No                 \\
\begin{tabular}[c]{@{}c@{}}Huber \textit{et al.} (2016)\\ \cite{huber2016young}\end{tabular}              & 4-6 Years                    & 50                       & Finger                            & No                 \\
\begin{tabular}[c]{@{}c@{}}Chen \textit{et al.} (2020)\\ \cite{chen2020examining}\end{tabular}              & 4-7 Years                    & 28                       & Finger                            & No                 \\
\begin{tabular}[c]{@{}c@{}}Woodward \textit{et al.} (2016)\\ \cite{woodward2016characterizing}\end{tabular}           & 5-10 Years                   & 30                       & Finger                                & No                 \\
\begin{tabular}[c]{@{}c@{}}Shaw and Anthony (2016)\\ \cite{shaw2016analyzing}\end{tabular}           & 5-10 Years                   & 24                       & Finger                                & No                 \\
\begin{tabular}[c]{@{}c@{}}Nacher \textit{et al.} (2018)\\ \cite{nacher2018examining}\end{tabular}             & 5-10 Years                   & 55                       & Finger                                 & No                 \\
\begin{tabular}[c]{@{}c@{}}Tabatabaey-Mashadi \textit{et al.} (2015)\\ \cite{tabatabaey2015analyses}\end{tabular} & 6-7 Years                    & 178                      & Stylus                                & No                 \\
\begin{tabular}[c]{@{}c@{}}Anthony \textit{et al.} (2014)\\ \cite{anthony2014designing}\end{tabular}            & 6-17 Years                   & 44                       & Finger                             & No                 \\
\begin{tabular}[c]{@{}c@{}}McKnight and Cassidy (2010)\\ \cite{mcknight2010children}\end{tabular}      & 7-10 Years                   & 80                       & Finger/Stylus                       & No                 \\
\begin{tabular}[c]{@{}c@{}}Arif and Sylla (2013)\\ \cite{arif2013comparative}\end{tabular}      & 8-11 Years                   & 12                       & Finger/Stylus               & No                              \\
\begin{tabular}[c]{@{}c@{}}Laniel \textit{et al.} (2020)\\ \cite{laniel2020kinematic}\end{tabular}             & 8-11 Years                   & 25                       & Stylus                             & No                 \\
\begin{tabular}[c]{@{}c@{}}Anthony \textit{et al.} (2016)\\ \cite{anthony2016gestures}\end{tabular}            & \textless 12 Years           & 24                       & Finger                                 & No                 \\ \hline
\textbf{ChildCIdb (Present Study)}                                                & \textbf{18 Months - 8 Years} & \textbf{438}             & \textbf{Finger/Stylus}       & \textbf{Yes} \\  \hline    
\end{tabular}
}
\label{table_relatedWorks}
\end{table*}

\section{Related Works}\label{related_works}
Different studies have evaluated the interaction of the children with mobile devices. Table~\ref{table_relatedWorks} shows a comparison of the most relevant studies in the literature ordered by the age of the subjects, including information such as the number of children considered in the study, the type of acquisition tool, etc. 

The first thing we would like to highlight is the lack of publicly available datasets in the field. To the best of our knowledge, the novel dataset presented in this study (ChildCIdb) is the only available dataset to date together with the dataset presented by Vatavu \textit{et al.}~\cite{vatavu2015touch}. Therefore, ChildCIdb is one of the main contributions of this study, not only due to the large number of children captured (438), but also to other many aspects such as the age of the children (from 18 months to 8 years), acquisition tool (touch and stylus), ADHD information, etc.

Analysing the studies focused on the first stage of the Piaget's theory (Sensorimotor, 0-2 years), to the best of our knowledge the work presented by Crescenzi and Grané~\cite{lanna2019touch} is the only one available. This is mainly produced due to the difficulties when capturing data from children in that age range (e.g., they sometimes do not want to play with the mobile devices). The focus of their study was to analyse how children under 3 years old interact with mobile devices, using commercial apps related to drawing and colouring tasks. They concluded that children of those ages almost exclusively employ the stroke (swipe) gesture to start interaction with the coloring app. Other gestures such as press (before 20 months) and tap (mostly after 24 months) are used in the drawing activity. Finally, they discovered that children under 3 adapt their gestures to the content of the apps and suggested that the use of app tools (e.g., color palette) may begin from 2 years. 

Many studies have focused on the second stage of the Piaget's theory (Preoperational, 2-7 years), paying special attention to the ability to perform gestures on multi-touch surfaces. Nacher \textit{et al.} proposed in~\cite{nacher2015multi} a set of 8 different tasks to measure the ability of the children to perform gestures. They concluded that children in the age 2-3 are able to perform simple gestures such as tap and drag and drop but also other complex ones such as one-finger rotation. However, some issues might appear while performing more complex gestures such as double tap, long press, scale down, and two-finger rotation. A similar research line was studied by Hiniker \textit{et al.}~\cite{hiniker2015touchscreen}, reviewing 100 touchscreen apps for preschoolers. In addition, the authors found that children above 3 are able to follow in-app audio instructions and on-screen demonstrations. 

An interesting article in this line is the work presented by Vatavu \textit{et al.}~\cite{vatavu2015touch}. In that work the authors captured and released to the research community a dataset composed of 89 children (3-6 years) and 30 young adults. They analysed the way children interact with mobile devices, showing significant improvements in children's touch performance as they grow from 3 to 6 years compared with the findings obtained by Nacher \textit{et al.}~\cite{nacher2015multi}. For example, children were able to perform gestures such as double tap and single-touch drag and drop. However, it seems that there are still some gesture limitations, e.g., the completion rate of tasks based on multi-touch drag and drop gestures was very low (53.7\%), and even lower (35\%) in other studies~\cite{mansor2008little}. As a result, the authors proposed different guidelines for designing children applications. Similar conclusions have been obtained in other studies in the literature~\cite{aziz2013children,hussain2016children}, and projects such as the Mobile Touch and Gesture Interaction for Children (MTAGIC)~\cite{anthony2019physical}. 

A very interesting study in this research line was presented by Chen \textit{et al.}~\cite{chen2020examining}. The aim of the study was to investigate how cognitive and motor development was related to children's touchscreen interaction. Experiments were carried out using 28 children in the ages 4-7, concluding that factors such as age, grade level, motor skill, and executive function show similar correlations with target miss and gesture recognition rates.

Mobile devices have also been studied as a way to teach children, in particular through digital game-based learning (DGBL) applications. Behnamnia \textit{et al.}~\cite{BEHNAMNIA2020105227} investigated whether DGBL can improve the creativity skills in preschool children (3-6 years). Nine different games were considered in the study, concluding that DGBL can potentially affect children's ability to develop creative skills and critical thinking, knowledge transfer, acquisition of skills in digital experience, and a positive attitude toward learning. Similar conclusions were extracted by Huber \textit{et al.}~\cite{huber2016young} when asking children to solve puzzle games.

Considering that children and adults typically use different interaction patterns on mobile devices, some studies have proposed the development of automatic systems to detect age groups. This research line has many different potential applications, e.g., restrict the access to adult contents or services such as on-line shopping. Shaw and Anthony~\cite{shaw2016analyzing} presented an analysis of gestures using 24 children (5-10 years) and 27 adults, considering features based on geometric, kinematic, and relative articulation. The authors discussed how children's gesturing abilities and behaviors differ between age groups, and from adults. Vera-Rodriguez \textit{et al.}~\cite{2019_LognormalityChapter_Vera} presented an automatic system able to detect children from adults with classification rates over 96\%. This detection system is based on the combination of features based on neuromotor skills, task time, and accuracy. The dataset released by Vatavu \textit{et al.}~\cite{vatavu2015touch} was considered in the experimental framework. In a related work, Acien \textit{et al.} proposed an enhanced detection system including global features from touch interaction~\cite{2018_IETB_DetectChildTouch_Acien}.

Not only the screen interaction using the finger has been studied as a way to interact with mobile devices. Different studies have considered the stylus for the acquisition tool. Remi \textit{et al.}~\cite{remi2015exploring} studied the scribbling activities executed by children of 3-6 years. They considered the Sigma-Lognormal writing generation model~\cite{2019_LognormalityChapter_Vera,o2009development} to analyse the motor skills, concluding that there are significant differences in the model parameters between ages. Stylus has also been considered by Tabatabaey-Mashadi \textit{et al.}~\cite{tabatabaey2015analyses} to analyse the correlation between the performance of polygonal shape drawing and the levels in handwriting performance. The study revealed that there are details in the children's drawing strategy highly related to the handwriting performance. Recently, Laniel \textit{et al.}~\cite{laniel2020kinematic} proposed a new measure of fine motor skills, the Pen Stroke Test (PST) in order to discriminate between children with and without attention-deficit/hyperactivity disorder (ADHD). This test is also based on the parameters of the Sigma-Lognormal model, providing preliminary evidences that the PST may be very useful for detecting ADHD.

In addition, children's interaction with both finger and stylus acquisition tools have been preliminary compared in the literature. McKnight and Cassidy~\cite{mcknight2010children} performed a comparison between pen stylus and finger in a controlled scenario considering children aged 7-10, providing a set of general guidelines for the design of mobile devices for children. Also, Arif and Sylla~\cite{arif2013comparative} performed a comparative evaluation of touch and pen gestures for adults and children in the age range of 8-11 years old. Results showed that gestures performed using the pen were significantly faster and more accurate than touch for adult users. However, no significant effect regarding pen and finger was observed on performance for children in those ages.  		

Finally, for completeness, we also include in Table~\ref{table_relatedWorks} the description of our ChildCIdb database presented in this study, which is planned to be extended yearly to enable longitudinal studies. In addition to the information related to finger/stylus interaction with age that allows to study the cognitive and neuromotor development of children, our ChildCI framework is also extended with other interesting children information such as emotion, the presence of ADHD, and the grades at the school. Our idea is to acquire and provide a rich database to the research community in order to further advance in other multiple research lines such as the relationship between children device interaction with emotions~\cite{voeffray2011emotion}, and with school grades~\cite{lee2009students}, among others. An interesting study in this line was presented by Sanches \textit{et al.} in~\cite{sanches2019hci}, extracting 139 papers on depression, anxiety, and bipolar health issues from 10 years of SIGCHI conference proceedings. Although most studies are focused on adults, some of them consider CCI scenarios proving the possible benefits of touch applications for children, e.g., to promote calmness in children with autism.

%All parents provided informed consent for their child'sparticipation, approved by the host institution's research ethicscommittee.

\begin{table*}[t]
\centering
\caption{Statistics of the ChildCIdb dataset regarding the number of children associated to each educational level, and the gender and handedness information.}
\scalebox{0.88}{
\begin{tabular}{cccccccc}
\multirow{2}{*}{\textbf{Educational Level}} & \multirow{2}{*}{\textbf{\# Subjects}} & \multicolumn{2}{c}{\textbf{Gender}} & \multicolumn{4}{c}{\textbf{Handedness}}                           \\ \cline{3-8} 
                                            &                                       & \textbf{Male}   & \textbf{Female}   & \textbf{Right} & \textbf{Left} & \textbf{Both} & \textbf{Unknown} \\ \hline
2 (1-2 Years)                               & 18                                    & 8               & 10                & 12             & 3             & 2             & 1                \\
3 (2-3 Years)                               & 36                                    & 14              & 22                & 30             & 3             & 3             & 0                \\
4 (3-4 Years)                               & 50                                    & 29              & 21                & 38             & 5             & 7             & 0                \\
5 (4-5 Years)                               & 66                                    & 32              & 34                & 58             & 6             & 1             & 1                \\
6 (5-6 Years)                               & 93                                    & 53              & 40                & 83             & 8             & 0             & 2                \\
7 (6-7 Years)                               & 77                                    & 35              & 42                & 69             & 8             & 0             & 0                \\
8 (7-8 Years)                               & 98                                    & 48              & 50                & 79             & 15            & 0             & 0                \\ \hline
\textbf{Total}                              & \textbf{438}                          & \textbf{219}    & \textbf{219}      & \textbf{369}   & \textbf{48}   & \textbf{17}   & \textbf{4}      
\end{tabular}
}
\label{tab:acquisition_statistics}
\end{table*}

\section{ChildCIdb Description}\label{sec:ChildD}

\subsection{Ethical Issues}\label{sec:ethical_issues}
There are potential ethical implications associated with two aspects of the study: \textit{i)} the participation of children under 8 years old, and \textit{ii)} the data acquisition and release to the research community, always keeping the identity of the subjects anonymous. In light of this, all parents of children participating in this study provided informed, written consent for their child's participation and data acquisition and release. In addition, and in order to acquire a large and longitudinal database, we follow two methods in a preliminary stage to explain all the details of ChildCI framework to the families: \textit{i)} face-to-face meetings in the school, allowing parents to play with the software application and ask all concerns regarding the research project, and \textit{ii)} sending via email an information sheet with the objectives of the study, procedure, privacy, etc. These information procedures had a very positive effect in the project, increasing transparency and participation of the families. In addition, and in order to keep the interest in the project (longitudinal study), the research advancements achieved are presented to the families and members of the school every year, with positive feedback from all of them.

\subsection{Acquisition: Year 1}\label{sec:enrolment}
ChildCIdb is a novel Child-Computer Interaction dataset. This is an on-going dataset collected in collaboration with the school GSD Las Suertes in Madrid, Spain. This article presents the first version of ChildCI dataset\footnote{\url{https://github.com/BiDAlab/ChildCIdb_v1}} (ChildCIdb v1), which comprises one capturing session with 438 children in total in the ages from 18 months to 8 years, grouped in 8 different educational levels according to the Spanish education system. Table~\ref{tab:acquisition_statistics} provides the statistics of ChildCIdb regarding the number of children associated to each educational level, and also the gender and handedness information. As can be seen, the number of children captured increases with the educational level, being levels 2 and 3 the levels with less subjects. As commented before, this is produced due to: \textit{i)} less children are grouped in the same class, and \textit{ii)} the acquisition is usually more difficult as they are very young. Regarding the gender statistics of the ChildCIdb, 50\% of the children were male/female whereas for the handedness, 84\% were right-handed, although this factor is not completely defined until they are 5 years old.

In addition to the gender and handedness information, the following personal information was acquired during the enrolment stage through the informed, written consent: \textit{i)} date of birth, and if he/she is premature (gestation period of less than 37 weeks), \textit{ii)} whether he/she is a child with ADHD, \textit{iii)} whether he/she has ever used any mobile device before the acquisition, and \textit{iv)} his/her educational grades. All this information enriches the project, being possible to research in several interesting lines, e.g., is there any relationship between the way children interact with the devices and their grades? 

Finally, as most of the children of the study had already experience with mobile devices according to the information provided by the families, we do not consider a familiarisation phase in our framework. No help was provided to the children apart from the instructions indicated on the screen before the beginning of each test. For children under 3 years old, oral instructions were also given following the conclusions extracted in~\cite{hiniker2015touchscreen}. As this is a longitudinal study composed of several acquisition sessions (see Sec.~\ref{sec:yearly_acquisition}), we plan to investigate the interaction evolution of the children with time, and also regarding the previous experience of the children with mobile devices.

\begin{table}[t]
\centering
\caption{Statistics of the emotional analysis per educational level for the ChildCIdb. DK/DA stands for ``does not know/does not answer''.}
\scalebox{0.88}{
\begin{tabular}{ccccc}
\textbf{Educational Level} & \textbf{Happy} & \textbf{Normal} & \textbf{Sad} & \textbf{DK/DA} \\ \hline
2 (1-2 Years)              & 3              & 3               & 1            & 11             \\
3 (2-3 Years)              & 19             & 1               & 7            & 9              \\
4 (3-4 Years)              & 39             & 0               & 2            & 9              \\
5 (4-5 Years)              & 52             & 2               & 0            & 12             \\
6 (5-6 Years)              & 83             & 1               & 1            & 8              \\
7 (6-7 Years)              & 63             & 2               & 6            & 6              \\
8 (7-8 Years)              & 83             & 4               & 0            & 11             \\ \hline
\textbf{Total}             & \textbf{342}   & \textbf{13}     & \textbf{17}  & \textbf{66}   
\end{tabular}
}
\label{tab:emotion_statistics}
\end{table}

\subsection{Yearly Acquisition Plan}\label{sec:yearly_acquisition}
ChildCIdb is planned to be extended yearly to enable longitudinal studies. The same children considered in ChildCIdb v1 will be acquired as they grow up and move to the different educational levels (from 18 months to 8 years). Therefore, future versions of ChildCIdb will be extended to: \textit{i)} new children that are registered to the educational level 2 of the school GSD Las Suertes in Madrid, Spain; and \textit{ii)} new acquisition sessions for the children already captured in previous versions of ChildCIdb (up to 8 years old). The number of acquisition sessions and time gap between them will be different depending on the age of the children. For children between 1-4 years, we plan to have an acquisition every three months whereas for children between 4-8 years, acquisitions will take place every six months. This is motivated due to the quick motor and cognitive development changes suffered at early ages. Also, to enable longitudinal studies, the acquisition of ChildCIdb will last over 5 years. Finally, future acquisitions of ChildCIdb will implement a randomized strategy of the tests (at block level) to avoid possible learning effects. These aspects have been approved by neurologists, child psychologists, and educators of the GSD school.

\begin{figure*}[!]
\begin{center}
   \includegraphics[width=\linewidth]{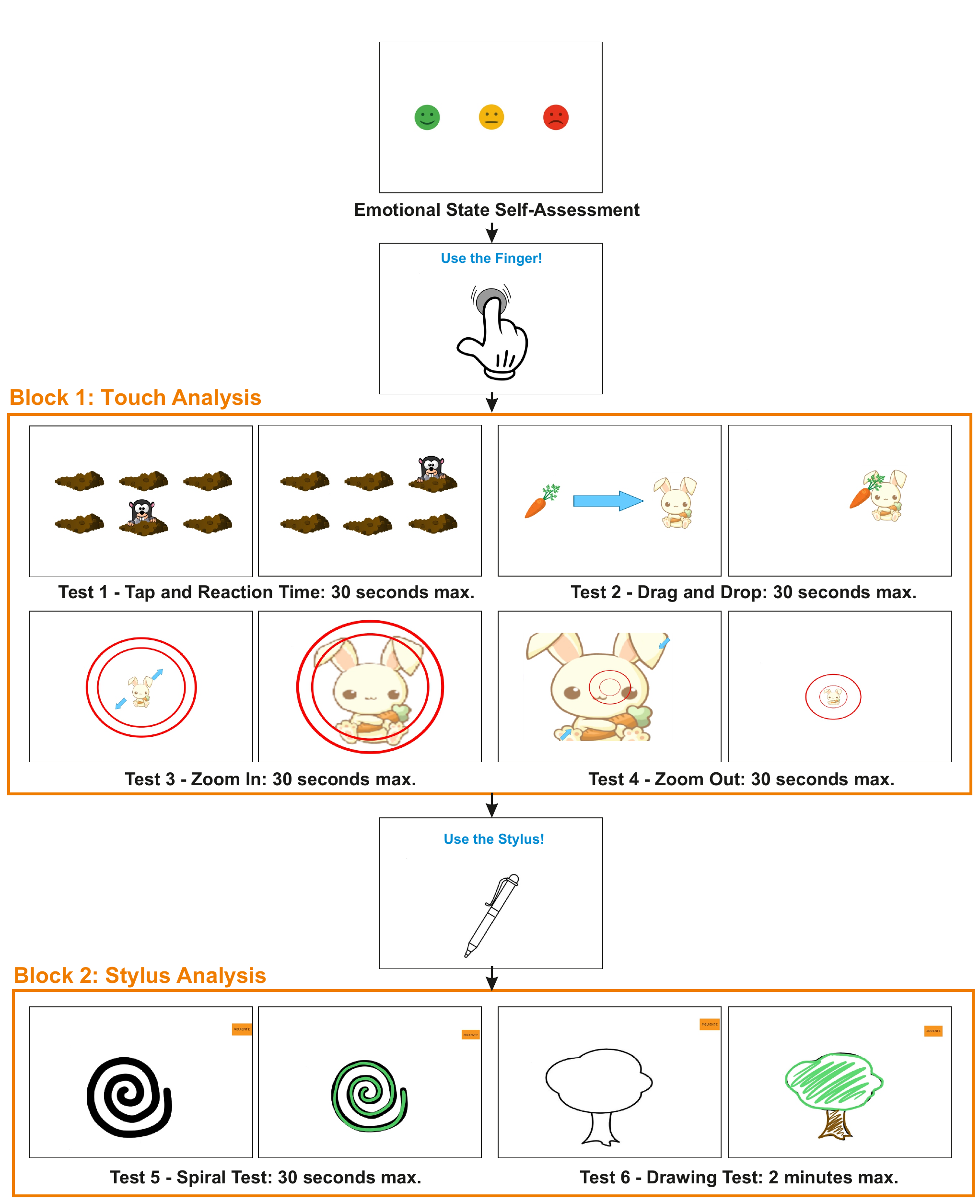}
\end{center}
   \caption{Examples of the different interfaces designed in ChildCI for each test, before and after their execution, including the maximum time set up in each of them. Two main acquisition blocks are considered: \textit{i)} touch, and \textit{ii)} stylus. Representative video recordings of the different educational levels are available at \url{https://github.com/BiDAlab/ChildCIdb_v1}
}
\label{fig:acquisition_application}
\end{figure*}

\subsection{Software Application}\label{sec:application}
An Android mobile application was implemented for the acquisition, which comprises 6 different tests grouped in 2 main blocks: \textit{i)} touch, and \textit{ii)} stylus. Tests were designed considering many of the cognitive and neuromuscular aspects highlighted in the state of the art, e.g., the evolution of children gestures with age. In addition, all tests were discussed and approved by neurologists, child psychologists, and educators of the GSD school. We describe now the procedure followed: \textit{i)} members of BiDA-Lab discussed first with the neurologists the theoretical aspects of the preliminary tests (test changes were considered at this stage based on the discussions, e.g., inclusion of Test 6 – Drawing Test); \textit{ii)} members of BiDA-Lab, together with neurologists, discussed the acceptance of the proposed tests with the child psychologists and educators of the GSD school. Aspects such as the maximum time of the test, age of the children, and type of activity to perform were considered in the selection of the final tests. Finally, a pilot study was carried out before the acquisition of ChildCIdb in order to confirm the appropriate designing of the software using children from different ages. The following designing aspects were validated in the pilot study: \textit{i)} the maximum time dedicated to each test and all the acquisition process, \textit{ii)} the number of tests included in the whole acquisition, \textit{iii)} the instructions provided to the children in each test, and \textit{iv)} the correct appearance of the content to call the attention of the children and motivate them.

Fig.~\ref{fig:acquisition_application} shows some examples of the different interfaces designed in ChildCI for each test, before and after their execution. As the participants are children, and keeping in mind they might not be able to focus on a task for a long time, we decided to develop a brief and interactive acquisition App in order to keep their attention as much as possible in a limited amount of time. Thus, we decided to set up a maximum time for each test as indicated in Fig.~\ref{fig:acquisition_application}, being 5 minutes the maximum time for the complete acquisition. In case the child is not able to finish each test in the maximum time set for it, the application automatically moves to the next test.

We first capture the emotional state of the children before the beginning of the acquisition. This meta-data information might be interesting for the project to answer, for example, the following question: is it possible to predict the children mood through the interaction with mobile devices? In our mobile application, three faces with different colours and facial expressions were represented on the screen, asking the children to touch one according to their emotional state. This is a simplification of a very complex attribute and might not reflect the real emotional state of the children. Table~\ref{tab:emotion_statistics} shows the statistics of the emotional state analysis per educational level for the first version of ChildCIdb. As can be seen, most children were in a good mood before the beginning of the acquisition (78\%). It is also interesting to remark the high number of children between 1-3 years that did not provide any information about their emotional state (DK/DA). As discussed in~\cite{lanna2019touch}, some children at the age of 3 can correctly label and recognise some emotions, as well as identify them in different situations. Nonetheless, we cannot assume that all children under 3 are conscious about emotional states. As a result, this emotional information should be interpreted carefully (specially for the youngest children).

The first block is focused on the analysis of the children motor and cognitive skills using their own finger as a tool. This block is indicated to the children through an image example. It comprises 4 different tests with different levels of difficulty to see the ability of the children to perform different hand gestures and movements. The maximum time of each test is 30 seconds. We describe next each of the tests:

\begin{itemize}
\item \textbf{Test 1 - Tap and Reaction Time:} the goal is to touch one mole at a time in order to see the ability of the children to perform tap gestures (gross motor skills) and their reaction times. Once the mole is touched, it disappears from that position and appears in another position of the screen. In total, 4 different moles must be touched for the end of the test. Just a single finger is needed to complete the task. 
\item \textbf{Test 2 - Drag and Drop:} the goal is to touch the carrot and swipe it to the rabbit. This test is designed to see the ability of the children to perform drag and drop gestures (fine motor skills). In order to facilitate the comprehension of the test and motivate the children, an intermittent blue arrow is shown in the screen until the children touch the carrot. Just a single finger is needed to complete the task. 
\item \textbf{Test 3 - Zoom In:} the goal is to enlarge the rabbit and put it inside the two red circles for a short time. This test is designed to: \textit{i)} analyse the ability of the children to perform scale-up (zoom-in) gestures, and \textit{ii)} analyse the precision of the motor control of the children when trying to put the rabbit inside the two red circles (fine motor skills). In order to facilitate the comprehension of the test, two intermittent outer arrows are depicted until the children touch the surface close to the rabbit. The rabbit can be only enlarged/shortened using two fingers. No displacement of the rabbit along the screen is allowed to facilitate the test, only the size of the rabbit can be modified.
\item \textbf{Test 4 - Zoom Out:} the goal of this test is similar to Test 3 but in this case children have to perform scale-down (zoom-out) gestures. Two fingers are needed to complete the test (fine motor skills).
\end{itemize}

After the end of the first block related to the children touch analysis, it starts the second block aimed to analyse the ability of the children using the pen stylus. This is indicated to the children through an image example showing a pen stylus. This block comprises the following 2 tests:

\begin{itemize}
\item \textbf{Test 5 - Spiral Test:} the goal of this test is to go across the spiral, from the inner part to the outer part, trying to keep it always in the area remarked in black colour. Once the children finish the test, they must press the button ``Next'' to move to the following test. The maximum timer set up for this test is 30 seconds. A similar version of this test is widely used for the detection of Parkinson's disease and movement disorders~\cite{fahn1993clinical}.
\item \textbf{Test 6 - Drawing Test:} the goal of this test is to colour the tree in the best way possible. Once the children decided to finish the test, they must press the button ``Next''. This last test ends the acquisition. The maximum timer set up for this test is 2 minutes. 
\end{itemize}

These tests are designed to investigate the cognitive and neuromotor skills of the children while performing actions with their own fingers or using the pen stylus, and also analyse their evolution with time. The research results that can be obtained by analysing ChildCI will be very valuable to better understand the current skills of the children in this society dominated by mobile devices.

\subsection{Acquisition Protocol}\label{sec:enrolment}
Currently, ChildCIdb comprises one acquisition session. The following principles were applied for the acquisition of the data:

\begin{itemize}
\item The same tablet device (Samsung Galaxy Tab A 10.1) was considered during all the acquisition process in order to avoid inter-device problems, e.g., different sampling frequencies~\cite{eBioSign_journal}.

\item All children performed the same tests in the same order (from Test 1 to Test 6) regardless of their educational level. This will allow us to perform a fair evaluation of the children inside a specific educational level and also between different ones.

\item No help was provided to the children apart from the instructions indicated on the screen before the beginning of each test. For children under 3 years old, oral instructions were also given following the conclusions extracted in~\cite{hiniker2015touchscreen}.

\item Children performed each test by themselves, without any other help.

\begin{figure}[!]
\centering
\subfigure{\label{tsne_real}
\includegraphics[width=0.88\linewidth]{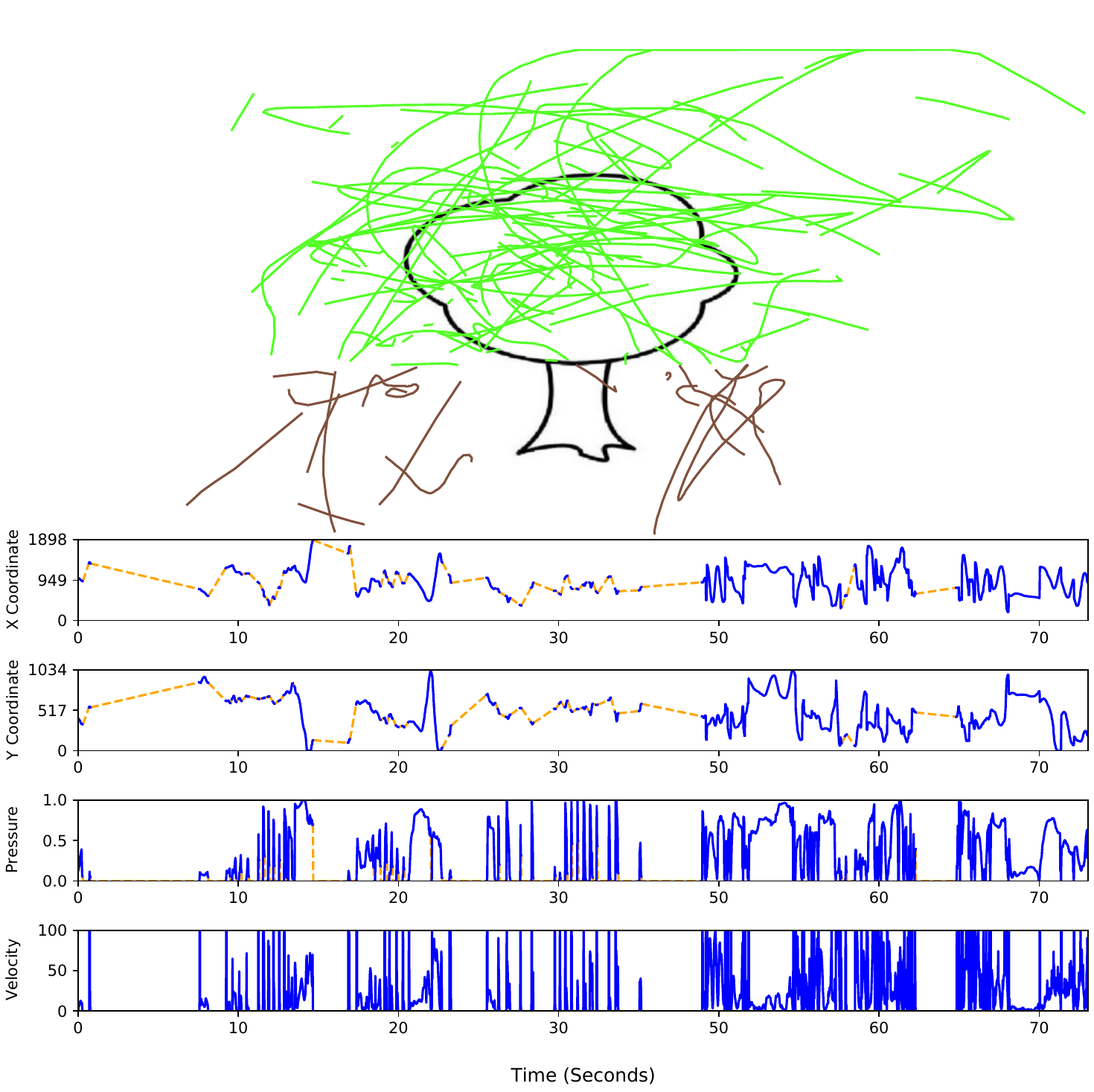}}
\subfigure{\label{tsne_synthetic}
\includegraphics[width=0.88\linewidth]{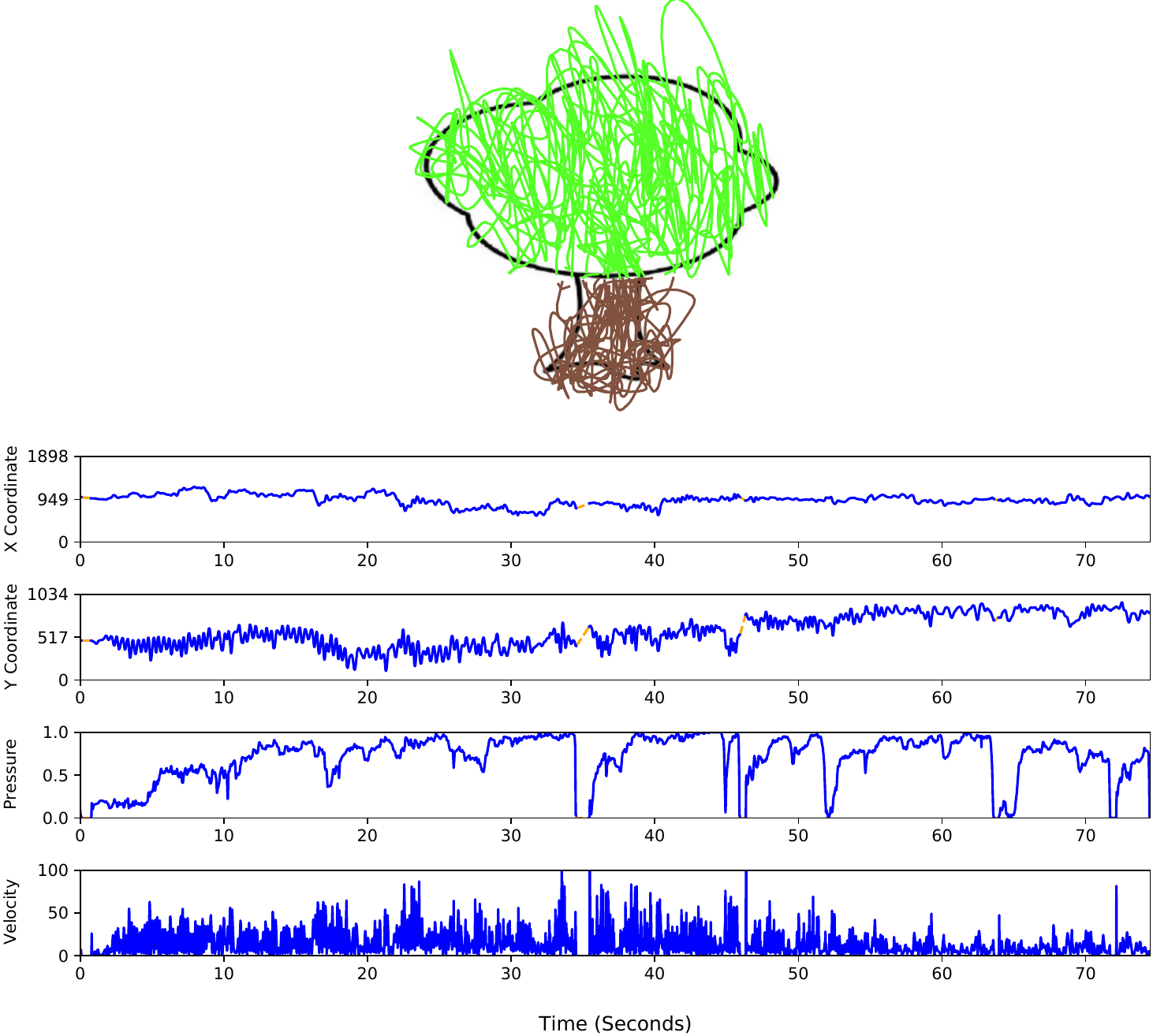}}
\subfigure{\label{tsne_synthetic}
\includegraphics[width=0.88\linewidth]{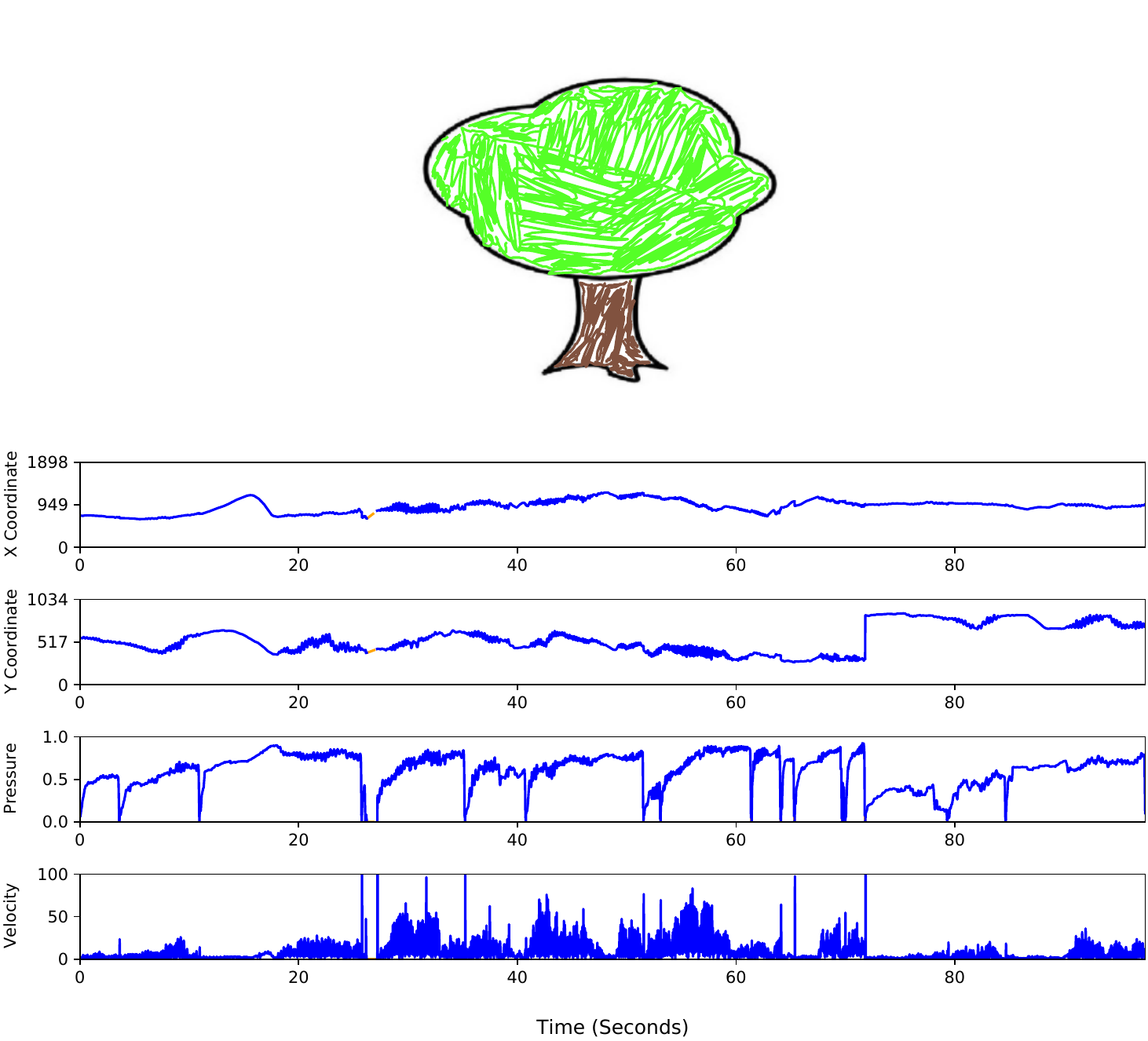}}
\caption{Examples of the Drawing Test performed by three different children age groups: (top) 1 to 3 years, (middle) 3 to 6 years, and (bottom) 6 to 8 years. Representative full video recordings of the different groups are available at \url{https://github.com/BiDAlab/ChildCIdb_v1}}
\label{fig:DT_examples}
\end{figure}

\item The acquisition was carried out inside the normal class, one at a time, and always with the child sitting far from the other children to avoid distractions, and with the device over a table. Children were allowed to move the device freely to feel comfortable.

\item The acquisition was controlled from a distance by a supervisor at all times in order to control the proper flow of the acquisition.

\end{itemize}

\section{Example Application: Age Detection}\label{sec:ChildD_applications}
This section analyses quantitatively one of the many different potential applications of ChildCIdb. In particular, we focus on the popular task of children age group detection based on the interaction with mobile devices~\cite{vatavu2015child,2019_LognormalityChapter_Vera,2018_IETB_DetectChildTouch_Acien}. Due to the large volume of information captured in ChildCIdb, we focus in this section only on the analysis of the Test 6 (Drawing Test) based on the way children colour a tree. Fig.~\ref{fig:DT_examples} shows some examples of the Drawing Test performed by different children age groups. 

The organisation of this section is as follows: Sec.~\ref{sec:experimental_protocol} describes the experimental protocol. Sec.~\ref{sec:detection_system} describes the age group detection systems proposed in this study. Finally, Sec.~\ref{sec:experimental_results} provides the results achieved.

\subsection{Experimental Protocol}\label{sec:experimental_protocol}
The experimental protocol proposed in this study has been designed to detect three different groups of children: Group 1 (children of educational levels 2 and 3, i.e., 1-3 years), Group 2 (children of educational levels 4, 5, and 6, i.e., 3-6 years), and finally Group 3 (children of educational levels 7 and 8, i.e., 6-8 years).

The current version of ChildCIdb (v1) has been divided into development (80\%) and evaluation (20\%) datasets, which comprise separate groups of subjects. The development dataset is used to train the age group detection systems whereas the evaluation dataset is finally used to test the trained systems on realistic conditions (new unseen subjects not used during the development stage). As the number of samples available in Group 1 and 3 is less than the Group 2, the data augmentation technique SMOTE of Imbalanced-Learn toolbox\footnote{\url{https://imbalanced-learn.org/stable/}} was considered only during the development stage to balance and better train the models. For the final evaluation, only real samples of ChildCIdb are considered. To better estimate the skill of the machine learning models proposed, \textit{k}-fold cross validation is used in this example application with \textit{k}=5. Final results provide the average values of the 5 fold cross validation.

\subsection{Age Group Detection Systems}\label{sec:detection_system}
Different machine learning approaches are studied in this work. The proposed age group detection systems comprise three main modules: feature extraction, feature selection, and classification. The specific parameters of each approach are selected over the development dataset.

\begin{table*}[t]
\centering
\caption{Novel set of 34 global features (denoted as Drawing features) proposed in this study for the task of colouring a tree (Test 6 - Drawing Test). $N$ stands for number and $T$ for time.}
\adjustbox{width=\textwidth}{
\begin{tabular}{|c|c|c|c|}
\hline
\textbf{\#} & \textbf{Feature Description}                                                         & \textbf{\#} & \textbf{Feature Description}                                                           \\ \hline
1  & $N$ (draw outside the tree margin)                       & 2  & $N$ (pen-downs)                                                                  \\ \hline
3  & $N$ (time samples inside the tree margin)                                                  & 4  & $N$ (time samples outside the tree margin)                                                   \\ \hline
5  & $N_{max}$ (pen-down time samples)                          & 6  & $T_{max}$ (pen-down)                                     \\ \hline
7  & $N_{min}$ (pen-down time samples)                                 & 8  & $T_{min}$ (pen-down)                                                                         \\ \hline
9  & $T_{mean}$ (pen-down)                                   & 10 & $N_{max}$ (pen-up time samples)                              \\ \hline
11 & $T_{max}$ (pen-up)                                                                         & 12 & $N_{min}$ (pen-up time samples)                                     \\ \hline
13 & $T_{min}$ (pen-up)                                     & 14 & $T_{mean}$ (pen-up)                                       \\ \hline
15 & $Mean$ (X-coordinate spatial position)                                                                & 16 & $Mean$ (Y-coordinate spatial position)                                                                                                                               \\ \hline
17 & $Std$ (X-coordinate spatial position)                                                                                                                    & 18 & $Std$ (Y-coordinate spatial position)                                                       \\ \hline
19 & $N$ (changes in drawing direction)                                                     & 20 & $Max$ (X-coordinate spatial position)                                                                          \\ \hline
21 & $Min$ (X-coordinate spatial position)                                                                        & 22 & $Max$ (Y-coordinate spatial position)                                                                          \\ \hline
23 & $Min$ (Y-coordinate spatial position)                                                                        & 24 & End test before time? (Yes/No)                                                         \\ \hline
25 & $T$ (drawing inside the tree margin)                                                & 26 & $T$ (drawing outside the tree margin)                                                 \\ \hline
27 & $T$ (drawing)                                                            & 28 & $T$ (not drawing)                                                  \\ \hline
29 & $T$ (drawing inside the tree margin) / $T$ (drawing) & 30 & $T$ (drawing outside the tree margin) / $T$ (drawing) \\ \hline
31 & $T$ (drawing inside the tree margin) / $T$ (drawing outside the tree margin)          & 32 & $T$ (drawing) / $T$ (Test)     \\ \hline
33 & Draw anything? (Yes/No)                                                    & 34 & $N$ (time samples)                                                                    \\ \hline
\end{tabular}
}
\label{tab:new_features}
\end{table*}

\subsubsection{\textbf{Feature Extraction}}\label{sec:feature_extraction} 
a set of 148 global features are extracted for each acquisition. From the total features extracted, 114 features are based on preliminary studies in the field of Human-Computer Interaction (HCI) and related with Time, Kinematic, Direction, Geometry, and Pressure information~\cite{Marcos08a,IWBF_2015_ForensicSignature}. The remainder 34 features (denoted as Drawing features) are originally presented in this study and designed for the specific Drawing Test (colouring a tree). Table~\ref{tab:new_features} describes this novel set of 34 global features, which extracts relevant information such as the length of the drawing strokes, and the number of times the children colour outside the margin of the tree, among many others.

\subsubsection{\textbf{Feature Selection}}\label{sec:feature_selection}
the following approaches are studied to select the most discriminative features from the total 148 global features originally extracted:

\begin{itemize}
\item \textit{Fisher Discriminant Ratio (FDR)}: it measures the discriminative power of each independent global feature. The value increases with the inter-class variability and decreases with the intra-class variability. In our experiments, we select the subset of global features whose FDR values are higher than 0.05.
\item \textit{Sequential Forward Floating Search (SFFS)}: this algorithm aims to select the optimal feature subset for a specific optimisation criteria while reducing the number of possible combinations to be tested. Therefore, this algorithm offers a suboptimal solution as it does not take into account all possible feature combinations, although it does consider correlations between features, achieving high-accuracy results~\cite{eBioSign_journal}. The specific implementation considered in this study is publicly available in MLxtend\footnote{\url{http://rasbt.github.io/mlxtend/}}.
\item \textit{Genetic Algorithm (GA)}: this algorithm is inspired by Charles Darwin's theory of natural evolution by relying on biologically inspired operations such as mutation, crossover, and selection. We consider the genetic algorithm originally presented in~\cite{10.5555/534133}. This algorithm has been completely programmed in this study from scratch, including aspects such as parallel execution to speed up the feature selection process. Our implemented code using Python is publicly available in GitHub\footnote{\url{https://github.com/BiDAlab/GeneticAlgorithm}}. In our experiments, we consider the following parameters: random generations = 100, population = 200, crossover rate = 0.6, mutation rate = 0.05. 
\end{itemize}

\subsubsection{\textbf{Classification}}\label{sec:classification}
different classifiers are studied in our example application. All of them are publicly available in Scikit-Learn\footnote{\url{https://scikit-learn.org/stable/}}. In addition, for each classifier, the optimal parameters are selected after an in-depth search over the development dataset using the class GridSearchCV of Scikit-Learn.

\begin{table*}[t]
\centering
\caption{Results achieved in terms of age group classification Accuracy (\%) over the final evaluation dataset of ChildCIdb for the different feature selection and classification approaches considered. We remark in bold the best result achieved.}
\adjustbox{width=0.9\textwidth}{
\begin{tabular}{c|ccccccc}
\textbf{Feature Selection} & \textbf{Naive Bayes} & \textbf{Logistic Regression} & \textbf{K-NN} & \textbf{Random Forest} & \textbf{AdaBoost} & \textbf{SVM}   & \textbf{MLP} \\ \hline
\textbf{FDR}               & 69.63                & 73.99                        & 71.24         & 75.56                  & 68.27             & 75.58          & 76.72        \\
\textbf{SFFS}              & 78.09                & 82.22                        & 81.98         & 88.69                  & 76.28             & \textbf{90.45} & 85.98        \\
\textbf{GA}                & 77.86                & 81.30                        & 77.86         & 80.37                  & 73.98             & 81.51          & 81.76       
\end{tabular}
}
\label{table_resultados}
\end{table*}

\begin{itemize}
\item \textit{Naive Bayes (NB):} this is a simple probabilistic classifier based on Bayes' theorem with the ``naive'' assumption of conditional independence between every pair of features given the value of the class variable. 
\item \textit{Logistic Regression (LR):} this is a statistical classifier that models the probability of a certain class using logistic functions. In our experiments, we consider L2 regularisation. 
\item \textit{K-Nearest Neighbours (K-NN):} this is a non-parametric method in which an event is assigned to the class most common among its \textit{k} nearest neighbours. In our experiments the number of neighbours is 5, and the algorithm used to compute the nearest neighbours is BallTree.
\item \textit{Random Forest (RF):} this is an ensemble learning method that fits a number of decision tree classifiers at training time and outputs the class that is the mode of the classes of the individual trees. In our experiments, the number of trees in the forest is 100, the maximum depth of the tree is 75, and the function to measure the quality of the split is gini.
\item \textit{AdaBoost (AB):} it combines multiple ``weak classifiers'' into a single ``strong classifier''. It begins by fitting a classifier on the original dataset and then fits additional copies of the classifier on the same dataset but where the weights of incorrectly classified instances are adjusted such that subsequent classifiers focus more on difficult cases. We consider here the AdaBoost-SAMME approach presented in~\cite{hastie2009multi} with 50 maximum number of estimators.
\item \textit{Support Vector Machines (SVM):} this is a popular learning algorithm that constructs a hyperplane or set of hyperplanes in a high- or infinite-dimensional space that best separates the classes. In this case, we have selected regularization with 0.1, polynomial kernel with degree 3 and coefficient scaled.  
\item \textit{Multi-Layer Perceptron (MLP):} this is a class of feedforward Artificial Neural Network (ANN). It consists of three or more layers (an input and an output layer with one or more hidden layers) of non-linear activation nodes. Each node is connected to every node in the following layer (fully-connected). In our study, we have considered four hidden layers with 100, 200, 200, and 100 neurons for each hidden layer, respectively. In addition, Adam optimiser is considered with default parameters (learning rate of 0.001) and a loss function
based on cross-entropy.
\end{itemize}

\begin{figure}[t]
\begin{center}
   \includegraphics[width=\linewidth]{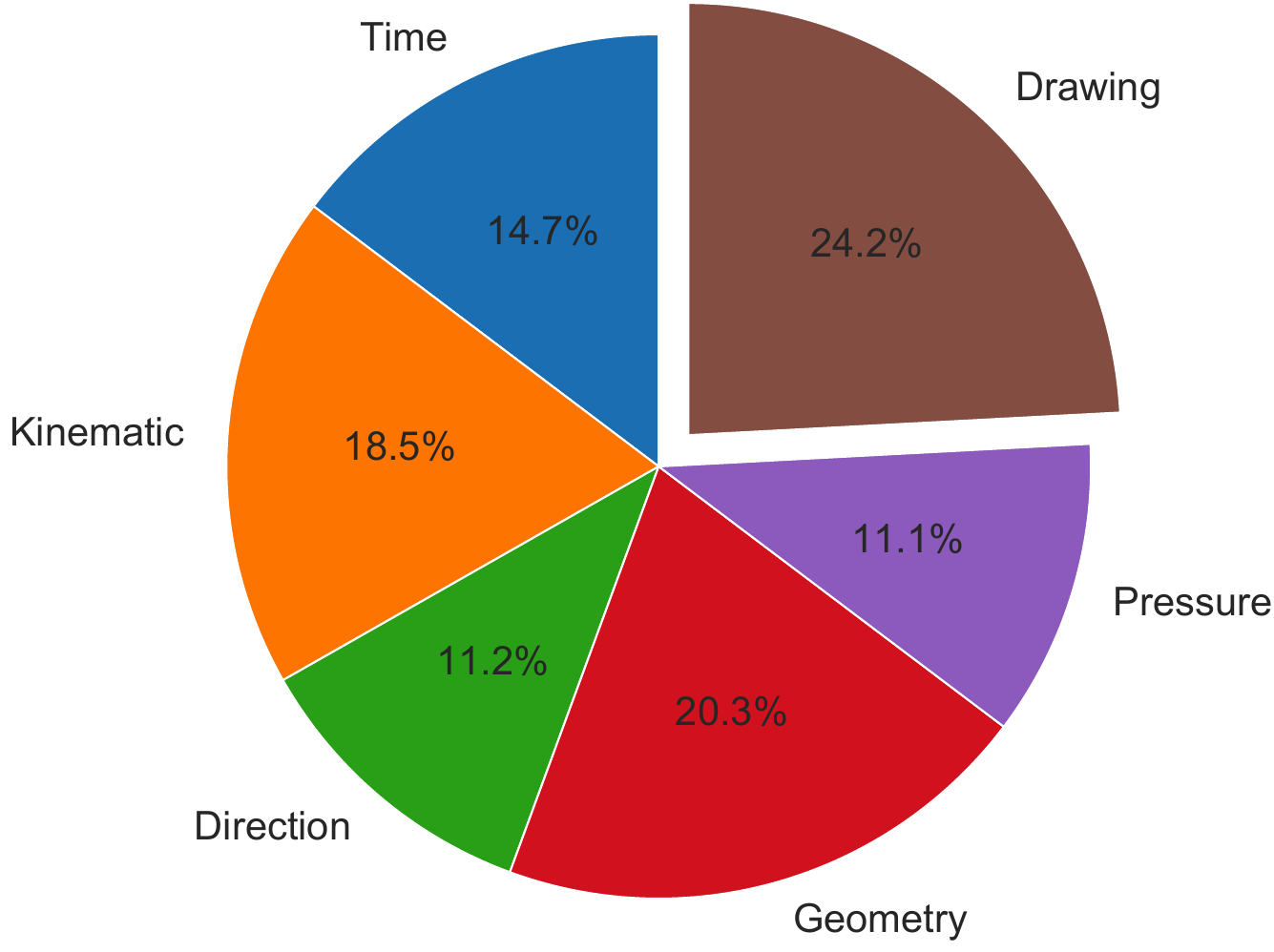}
\end{center}
   \caption{Average percentage of features selected per category.}
\label{fig:average_features}
\end{figure}

\subsection{Experimental Results}\label{sec:experimental_results}

\subsubsection{\textbf{Results}}\label{subsec_results}
Table~\ref{table_resultados} shows the results achieved in terms of age group classification Accuracy (\%) over the final evaluation dataset of ChildCIdb for the different feature selection and classification approaches considered.

We first analyse the results achieved by each feature selection technique. As can be seen, the algorithm SFFS provides the best results in all cases (83.38\% average accuracy), followed by the Genetic Algorithm (79.23\% average accuracy). The FDR algorithm provides the worst average accuracy results (73.00\%). This seems to be produced because the FDR feature selection technique is based on the discriminative power of each independent feature. No correlations between features are considered in the selection process.   

Analysing the results achieved by each classification approach, SVM, Random Forest, and MLP provide the best results with 90.45\%, 88.69\%, and 85.98\% accuracies, respectively. Other simpler classifiers such as Naive Bayes and K-NN provide much worse results (69.63\% and 71.24\% accuracies, respectively).

Finally, we compare the results achieved with the state of the art. To the best of our knowledge, this is the first study that focuses on the classification of children age groups (from 18 months to 8 years) based on the interaction with mobile devices. Previous studies were focused on a simpler task, i.e., classification between children (3-6 years) and adults~\cite{vatavu2015child,2019_LognormalityChapter_Vera,2018_IETB_DetectChildTouch_Acien}, achieving in the best cases classification accuracy results of 96.3\%. Comparing that result achieved in a simpler task with the results achieved in the present study (accuracy results over 90\%), we can conclude that: \textit{i)} good results are achieved, proving the soundness of the proposed age group classification systems, and \textit{ii)} the possibility to distinguish with high-accuracy results between different children age groups. 

\begin{figure}[t]
\begin{center}
   \includegraphics[width=0.95\linewidth]{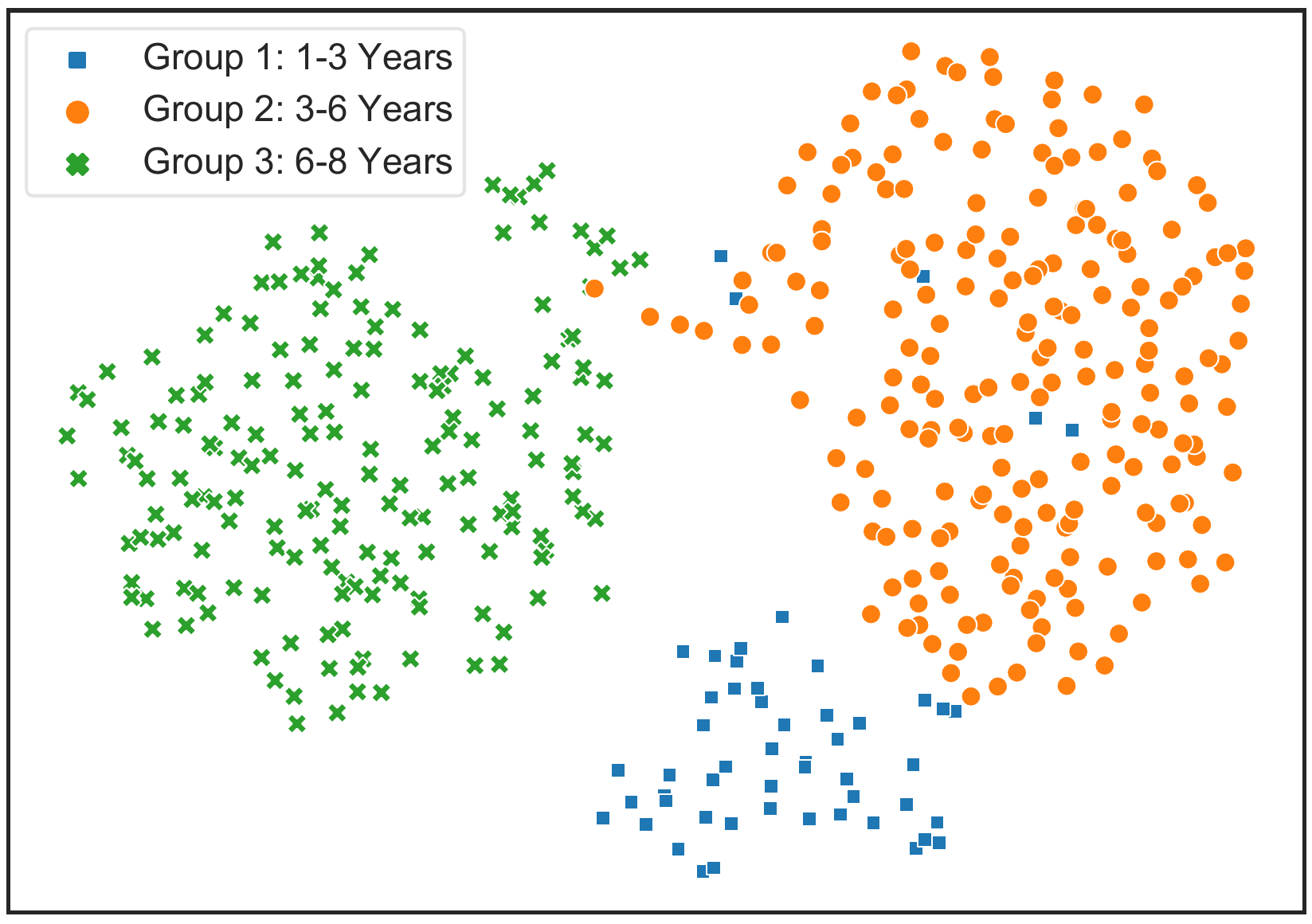}
\end{center}
   \caption{Projection of the best features selected in Sec.~\ref{subsec_results}, SFFS + SVM, into the 2D space generated using UMAP. Each point corresponds to one child of ChildCIdb v1.}
\label{fig:UMAP_features}
\end{figure}

\subsubsection{\textbf{Feature Analysis}}\label{subsec_feature_analysis}

this section analyses the type of features selected by the machine learning approaches studied in Sec.~\ref{subsec_results}. Fig.~\ref{fig:average_features} shows the average percentage of features selected per category, i.e., Time, Kinematic, Direction, Geometry, Pressure, and Drawing~\cite{Marcos08a}. In general, we can see that the novel features related to the Drawing information are the most selected ones with an average 24.2\%. This result proves the success of the novel features designed in this study for the task of children age group detection. Other features based on the Geometry (20.3\%) and Kinematic (18.5\%) information of the children while interacting with the devices are also very important to distinguish between different age groups. However, information related to the Direction and Pressure performed by the children while colouring the tree seems not to be very discriminative to distinguish between children age groups (11.2\% and 11.1\% respectively). 

Finally, for completeness, we apply a popular data visualisation method to show the feature distributions across age groups. In particular, we select the set of features that provides the best results in Sec.~\ref{subsec_results}, i.e., SFFS + SVM. Fig.~\ref{fig:UMAP_features} shows the unsupervised Uniform Manifold Approximation and Projection (UMAP)~\cite{mcinnes2018umap} projections for each of the children considered in ChildCIdb v1. We have coloured each point according to its age group groundtruth. As can be seen, the consequent feature representation results in three clusters highly correlated with the age groups. The age information is highly embedded in the feature vector and a simple unsupervised algorithm such as UMAP reveals the presence of this information.

These results prove the existence of different patterns in the motor control process of the children with the age such as the velocity and acceleration while performing strokes. These insights also agree with the physiological and cognitive changes across age discussed in Piaget's theory~\cite{piaget2008psychology}.

\section{Limitations and Future Work}\label{limitations}
Some aspects of the current ChildCI framework could be improved. In particular, the correlation of the information extracted from our study with other popular standard tests considered for children development such as the NIH toolbox\footnote{\url{https://www.nia.nih.gov/research/resource/nih-toolbox}}, the Bayley Scales of Infant and Toddler Development~\cite{albers2007test}, and the Mullen Scales of Early Learning~\cite{mullen1995mullen}. This aspect could further benefit the insights extracted in our framework, and at the same time improve those popular tests with more quantitative measures of the motor and cognitive evolution of the children. Also, in addition to the Piaget's theory popularly considered in the literature, more recent theories related to the children development process could provide a different point of view and insights in our framework~\cite{thomas2000comparing}. These aspects will be considered in future work.

In addition, future work will be oriented to: \textit{i)} extend ChildCIdb with more participants and acquisition sessions, \textit{ii)} analyse and improve the accuracy of the children age group detection systems using the remaining tests of ChildCIdb not considered in the present article, \textit{iii)} study the application of other feature and signal representations of the drawing and screen interaction beyond the ones tested here with special emphasis in recent deep learning methods~\cite{2021_AAAI_DeepWriteSYN}, \textit{iv)} develop child-independent interaction models for the different test from which child-dependent behaviours can be derived, \textit{v)} correlate the interaction information with the meta-data stored in the dataset like learning outcomes and ADHD, \textit{vi)} combine the information provided by the multiple tests using information fusion methods, and \textit{vii)} exploit ChildCIdb in other research problems around e-Learning~\cite{2020_COMPSAC_HRedBB_JHO}, and e-Health~\cite{2019_FG_PDhandw_Castrillon}.

\section{Conclusions}\label{conclusions}
This article has presented a preliminary study of our framework named ChildCI, which is aimed at generating a better understanding of Child-Computer Interactions with applications to e-Health and e-Learning, among others.

In particular, in this article we have presented all the details regarding the design and development of a new child mobile application, the specific acquisition protocol considered, and the first capturing session of the ChildCI dataset (ChildCIdb v1), which is publicly available for research purposes\footnote{\url{https://github.com/BiDAlab/ChildCIdb_v1}}. In the scenario considered, children interact with a tablet device, using both the pen stylus and also the finger, performing different tasks that require different levels of motor and cognitive skills. ChildCIdb v1 comprises over 400 children in the ages from 18 months to 8 years, considering therefore the first three stages of the motor and cognitive development of the Piaget's theory.

In addition, we have demonstrated the potential of ChildCIdb including experimental results for one of the many possible applications: children age group detection. Different machine learning approaches have been studied, proposing a new set of 34 global features to automatically detect the age group, achieving accuracy results over 90\% and interesting findings in terms of the type of features more useful.

\section*{Acknowledgements}
This work has been supported by projects: PRIMA (H2020-MSCA-ITN-2019-860315), TRESPASS-ETN (H2020-MSCA-ITN-2019-860813), INTER-ACTION (PID2021-126521OB-I00 MICINN/FEDER), and Orange Labs.  This is an on-going project carried out with the collaboration of the school GSD Las Suertes in Madrid, Spain.



%\bibliographystyle{IEEEtran}

%

{
\bibliographystyle{IEEEtran}
\bibliography{egbib2}

\begin{thebibliography}{10}
\providecommand{\url}[1]{#1}
\csname url@samestyle\endcsname
\providecommand{\newblock}{\relax}
\providecommand{\bibinfo}[2]{#2}
\providecommand{\BIBentrySTDinterwordspacing}{\spaceskip=0pt\relax}
\providecommand{\BIBentryALTinterwordstretchfactor}{4}
\providecommand{\BIBentryALTinterwordspacing}{\spaceskip=\fontdimen2\font plus
\BIBentryALTinterwordstretchfactor\fontdimen3\font minus
  \fontdimen4\font\relax}
\providecommand{\BIBforeignlanguage}[2]{{%
\expandafter\ifx\csname l@#1\endcsname\relax
\typeout{** WARNING: IEEEtran.bst: No hyphenation pattern has been}%
\typeout{** loaded for the language `#1'. Using the pattern for}%
\typeout{** the default language instead.}%
\else
\language=\csname l@#1\endcsname
\fi
#2}}
\providecommand{\BIBdecl}{\relax}
\BIBdecl

\bibitem{kilicc2019exposure}
A.~O. K{\i}l{\i}{\c{c}}, E.~Sari, H.~Yucel, M.~M. O{\u{g}}uz, E.~Polat, E.~A.
  Acoglu, and S.~Senel, ``{Exposure to and Use of Mobile Devices in Children
  Aged 1-60 Months},'' \emph{European Journal of Pediatrics}, vol. 178, no.~2,
  pp. 221--227, 2019.

\bibitem{kucirkova2020lessons}
N.~Kucirkova, C.~Evertsen-Stanghelle, I.~Studsr{\o}d, I.~B. Jensen, and
  I.~St{\o}rksen, ``{Lessons for Child-Computer Interaction Studies Following
  the Research Challenges During the COVID-19 Pandemic},'' \emph{International
  Journal of Child-Computer Interaction}, p. 100203, 2020.

\bibitem{tsvyatkova2019review}
D.~Tsvyatkova and C.~Storni, ``{A Review of Selected Methods, Techniques and
  Tools in Child-Computer Interaction (CCI) Developed/Adapted to Support
  Children’s Involvement in Technology Development},'' \emph{International
  Journal of Child-Computer Interaction}, vol.~22, p. 100148, 2019.

\bibitem{piaget2008psychology}
J.~Piaget and B.~Inhelder, \emph{{The Psychology of the Child}}.\hskip 1em plus
  0.5em minus 0.4em\relax Basic books, 2008.

\bibitem{vatavu2015child}
R.-D. Vatavu, L.~Anthony, and Q.~Brown, ``{Child or Adult? Inferring Smartphone
  Users’ Age Group from Touch Measurements Alone},'' in \emph{Proc.
  Conference on Human-Computer Interaction}, 2015.

\bibitem{anthony2016gestures}
L.~Anthony, K.~A. Stofer, A.~Luc, and J.~O. Wobbrock, ``{Gestures by Children
  and Adults on Touch Tables and Touch Walls in a Public Science Center},'' in
  \emph{Proc. International Conference on Interaction Design and Children},
  2016.

\bibitem{2019_LognormalityChapter_Vera}
R.~Vera-Rodriguez, R.~Tolosana, J.~Hernandez-Ortega, A.~Acien, A.~Morales,
  J.~Fierrez, and J.~Ortega-Garcia, ``{Modeling the Complexity of Signature and
  Touch-Screen Biometrics using the Lognormality Principle},'' in \emph{The
  Lognormality Principle and its Applications}, R.~Plamondon, A.~Marcelli, and
  M.~A. Ferrer, Eds.\hskip 1em plus 0.5em minus 0.4em\relax World Scientific,
  2020.

\bibitem{2018_IETB_DetectChildTouch_Acien}
A.~Acien, A.~Morales, J.~Fierrez, R.~Vera-Rodriguez, and J.~Hernandez-Ortega,
  ``{Active Detection of Age Groups Based on Touch Interaction},'' \emph{IET
  Biometrics}, vol.~8, no.~1, pp. 101--108, 2019.

\bibitem{mcknight2010children}
L.~McKnight and B.~Cassidy, ``{Children's Interaction with Mobile Touch-Screen
  Devices: Experiences and Guidelines for Design},'' \emph{International
  Journal of Mobile Human Computer Interaction}, vol.~2, no.~2, pp. 1--18,
  2010.

\bibitem{anthony2014designing}
L.~Anthony, Q.~Brown, B.~Tate, J.~Nias, R.~Brewer, and G.~Irwin, ``{Designing
  Smarter Touch-Based Interfaces for Educational Contexts},'' \emph{Personal
  and Ubiquitous Computing}, vol.~18, no.~6, pp. 1471--1483, 2014.

\bibitem{vatavu2015touch}
R.-D. Vatavu, G.~Cramariuc, and D.~M. Schipor, ``{Touch Interaction for
  Children Aged 3 to 6 Years: Experimental Findings and Relationship to Motor
  Skills},'' \emph{International Journal of Human-Computer Studies}, vol.~74,
  pp. 54--76, 2015.

\bibitem{lanna2019touch}
L.~C. Lanna and M.~G. Oro, ``{Touch Gesture Performed by Children Under 3 Years
  Old when Drawing and Coloring on a Tablet},'' \emph{International Journal of
  Human-Computer Studies}, vol. 124, pp. 1--12, 2019.

\bibitem{2020_COMPSAC_HRedBB_JHO}
J.~Hernandez-Ortega, R.~Daza, A.~Morales, J.~Fierrez, and R.~Tolosana, ``{Heart
  Rate Estimation from Face Videos for Student Assessment: Experiments on
  edBB},'' in \emph{Proc. IEEE Conference on Computers, Software, and
  Applications}, 2020.

\bibitem{2019_FG_PDhandw_Castrillon}
R.~Castrillon, A.~Acien, J.~R. Orozco-Arroyave, A.~Morales, J.~F. Vargas,
  R.~Vera-Rodriguez, J.~Fierrez, J.~Ortega-Garcia, and A.~Villegas,
  ``{Characterization of the Handwriting Skills as a Biomarker for Parkinson
  Disease},'' in \emph{Proc. IEEE Int. Conf. on Automatic Face and Gesture
  Recognition}, 2019.

\bibitem{nacher2015multi}
V.~Nacher, J.~Jaen, E.~Navarro, A.~Catala, and P.~Gonz{\'a}lez, ``{Multi-Touch
  Gestures for Pre-Kindergarten Children},'' \emph{International Journal of
  Human-Computer Studies}, vol.~73, pp. 37--51, 2015.

\bibitem{hiniker2015touchscreen}
A.~Hiniker, K.~Sobel, S.~R. Hong, H.~Suh, I.~Irish, D.~Kim, and J.~A. Kientz,
  ``{Touchscreen Prompts for Preschoolers: Designing Developmentally
  Appropriate Techniques for Teaching Young Children to Perform Gestures},'' in
  \emph{Proc. International Conference on Interaction Design and Children},
  2015.

\bibitem{aziz2013children}
N.~A.~A. Aziz, ``{Children’s Interaction with Tablet Applications: Gestures
  and Interface Design},'' \emph{Children}, vol.~2, no.~3, pp. 447--450, 2013.

\bibitem{remi2015exploring}
C.~R{\'e}mi, J.~Vaillant, R.~Plamondon, L.~Prevost, and T.~Duval, ``{Exploring
  the Kinematic Dimensions of Kindergarten Children's Scribbles},'' in
  \emph{Proc. Conference of the International Graphonomics Society}, 2015.

\bibitem{BEHNAMNIA2020105227}
N.~Behnamnia, A.~Kamsin, M.~A.~B. Ismail, and A.~Hayati, ``{The Effective
  Components of Creativity in Digital Game-Based Learning among Young Children:
  A Case Study},'' \emph{Children and Youth Services Review}, vol. 116, p.
  105227, 2020.

\bibitem{hussain2016children}
N.~H. Hussain, T.~S.~M. Tengku~Wook, S.~F. Mat~Noor, and H.~Mohamed,
  ``{Children's Interaction Ability Towards Multi-Touch Gestures},''
  \emph{International Journal on Advanced Science, Engineering and Information
  Technology}, vol.~6, no.~6, pp. 875--881, 2016.

\bibitem{huber2016young}
B.~Huber, J.~Tarasuik, M.~N. Antoniou, C.~Garrett, S.~J. Bowe, J.~Kaufman, and
  S.~B. Team, ``{Young Children's Transfer of Learning from a Touchscreen
  Device},'' \emph{Computers in Human Behavior}, vol.~56, pp. 56--64, 2016.

\bibitem{chen2020examining}
Z.~Chen, Y.-P. Chen, A.~Shaw, A.~Aloba, P.~Antonenko, J.~Ruiz, and L.~Anthony,
  ``{Examining the Link between Children's Cognitive Development and
  Touchscreen Interaction Patterns},'' in \emph{Proc. International Conference
  on Multimodal Interaction}, 2020.

\bibitem{woodward2016characterizing}
J.~Woodward, A.~Shaw, A.~Luc, B.~Craig, J.~Das, P.~Hall~Jr, A.~Holla, G.~Irwin,
  D.~Sikich, Q.~Brown \emph{et~al.}, ``{Characterizing How Interface Complexity
  Affects Children's Touchscreen Interactions},'' in \emph{Proc. Conference on
  Human Factors in Computing Systems}, 2016.

\bibitem{shaw2016analyzing}
A.~Shaw and L.~Anthony, ``{Analyzing the Articulation Features of Children's
  Touchscreen Gestures},'' in \emph{Proc. ACM International Conference on
  Multimodal Interaction}, 2016.

\bibitem{nacher2018examining}
V.~Nacher, D.~C{\'a}liz, J.~Jaen, and L.~Mart{\'\i}nez, ``{Examining the
  Usability of Touch Screen Gestures for Children with Down Syndrome},''
  \emph{Interacting with Computers}, vol.~30, no.~3, pp. 258--272, 2018.

\bibitem{tabatabaey2015analyses}
N.~Tabatabaey-Mashadi, R.~Sudirman, R.~M. Guest, and P.~I. Khalid, ``{Analyses
  of Pupils’ Polygonal Shape Drawing Strategy with Respect to Handwriting
  Performance},'' \emph{Pattern Analysis and Applications}, vol.~18, no.~3, pp.
  571--586, 2015.

\bibitem{arif2013comparative}
A.~S. Arif and C.~Sylla, ``{A Comparative Evaluation of Touch and Pen gestures
  for Adult and Child Users},'' in \emph{Proc. 12th International Conference on
  Interaction Design and Children}, 2013.

\bibitem{laniel2020kinematic}
P.~Laniel, N.~Faci, R.~Plamondon, M.~H. Beauchamp, and B.~Gauthier,
  ``{Kinematic Analysis of Fast Pen Strokes in Children with ADHD},''
  \emph{Applied Neuropsychology: Child}, vol.~9, no.~2, pp. 125--140, 2020.

\bibitem{mansor2008little}
E.~I. Mansor, A.~De~Angeli, and O.~De~Bruijn, ``{Little Fingers on the
  Tabletop: A Usability Evaluation in the Kindergarten},'' in \emph{Proc. IEEE
  International Workshop on Horizontal Interactive Human Computer Systems},
  2008.

\bibitem{anthony2019physical}
L.~Anthony, ``{Physical Dimensions of Children's Touchscreen Interactions:
  Lessons from Five Years of Study on the MTAGIC Project},''
  \emph{International Journal of Human-Computer Studies}, vol. 128, pp. 1--16,
  2019.

\bibitem{o2009development}
C.~O’Reilly and R.~Plamondon, ``{Development of a Sigma-Lognormal
  Representation for On-Line Signatures},'' \emph{Pattern Recognition},
  vol.~42, no.~12, pp. 3324--3337, 2009.

\bibitem{voeffray2011emotion}
S.~Voeffray, ``{Emotion-Sensitive Human-Computer Interaction (HCI): State of
  the Art-Seminar Paper},'' \emph{Emotion Recognition}, pp. 1--4, 2011.

\bibitem{lee2009students}
J.~Lee and H.~Spires, ``{What Students Think About Technology and Academic
  Engagement in School: Implications for Middle Grades Teaching and
  Learning},'' \emph{AACE Journal}, vol.~17, no.~2, pp. 61--81, 2009.

\bibitem{sanches2019hci}
P.~Sanches, A.~Janson, P.~Karpashevich, C.~Nadal, C.~Qu, C.~Daud{\'e}n~Roquet,
  M.~Umair, C.~Windlin, G.~Doherty, K.~H{\"o}{\"o}k \emph{et~al.}, ``{HCI and
  Affective Health: Taking Stock of a Decade of Studies and Charting Future
  Research Directions},'' in \emph{Proc. Conference on Human Factors in
  Computing Systems}, 2019.

\bibitem{fahn1993clinical}
S.~Fahn, E.~Tolosa, and C.~Mar{\'\i}n, ``{Clinical Rating Scale for Tremor},''
  \emph{Parkinson's Disease and Movement Disorders}, vol.~2, pp. 271--280,
  1993.

\bibitem{eBioSign_journal}
R.~Tolosana, R.~Vera-Rodriguez, J.~Fierrez, A.~Morales, and J.~Ortega-Garcia,
  ``{Benchmarking Desktop and Mobile Handwriting across COTS Devices: the
  e-BioSign Biometric Database},'' \emph{PLOS ONE}, 2017.

\bibitem{Marcos08a}
M.~Martinez-Diaz, J.~Fierrez, J.~Galbally, and J.~Ortega-Garcia, ``{Towards
  Mobile Authentication Using Dynamic Signature Verification: Useful Features
  and Performance Evaluation},'' in \emph{Proc. Int. Conf. on Pattern
  Recognition}, 2008.

\bibitem{IWBF_2015_ForensicSignature}
R.~Tolosana, R.~Vera-Rodriguez, J.~Fierrez, and J.~Ortega-Garcia,
  ``{Feature-Based Dynamic Signature Verification under Forensic Scenarios},''
  in \emph{Proc. International Workshop on Biometrics and Forensics}, 2015.

\bibitem{10.5555/534133}
D.~E. Goldberg, \emph{{Genetic Algorithms in Search, Optimization and Machine
  Learning}}.\hskip 1em plus 0.5em minus 0.4em\relax Addison-Wesley, 1989.

\bibitem{hastie2009multi}
T.~Hastie, S.~Rosset, J.~Zhu, and H.~Zou, ``{Multi-Class AdaBoost},''
  \emph{Statistics and Its Interface}, vol.~2, no.~3, pp. 349--360, 2009.

\bibitem{mcinnes2018umap}
L.~McInnes, J.~Healy, and J.~Melville, ``{UMAP: Uniform Manifold Approximation
  and Projection for Dimension Reduction},'' \emph{arXiv preprint
  arXiv:1802.03426}, 2018.

\bibitem{albers2007test}
C.~A. Albers and A.~J. Grieve, ``{Test Review: Bayley, N.(2006). Bayley Scales
  of Infant and Toddler Development-Third Edition. San Antonio, TX: Harcourt
  Assessment},'' \emph{Journal of Psychoeducational Assessment}, vol.~25,
  no.~2, pp. 180--190, 2007.

\bibitem{mullen1995mullen}
E.~M. Mullen \emph{et~al.}, \emph{{Mullen Scales of Early Learning}}.\hskip 1em
  plus 0.5em minus 0.4em\relax AGS Circle Pines, MN, 1995.

\bibitem{thomas2000comparing}
R.~M. Thomas, \emph{{Comparing Theories of Child Development}}.\hskip 1em plus
  0.5em minus 0.4em\relax Wadsworth/Thomson Learning, 2000.

\bibitem{2021_AAAI_DeepWriteSYN}
R.~Tolosana, P.~Delgado-Santos, A.~Perez-Uribe, R.~Vera-Rodriguez, J.~Fierrez,
  and A.~Morales, ``{DeepWriteSYN: On-Line Handwriting Synthesis via Deep
  Short-Term Representations},'' in \emph{Proc. 35th AAAI Conference on
  Artificial Intelligence}, 2021.

\end{thebibliography}
}

\begin{IEEEbiography}[{\includegraphics[width=1in,height=1.25in,clip,keepaspectratio]{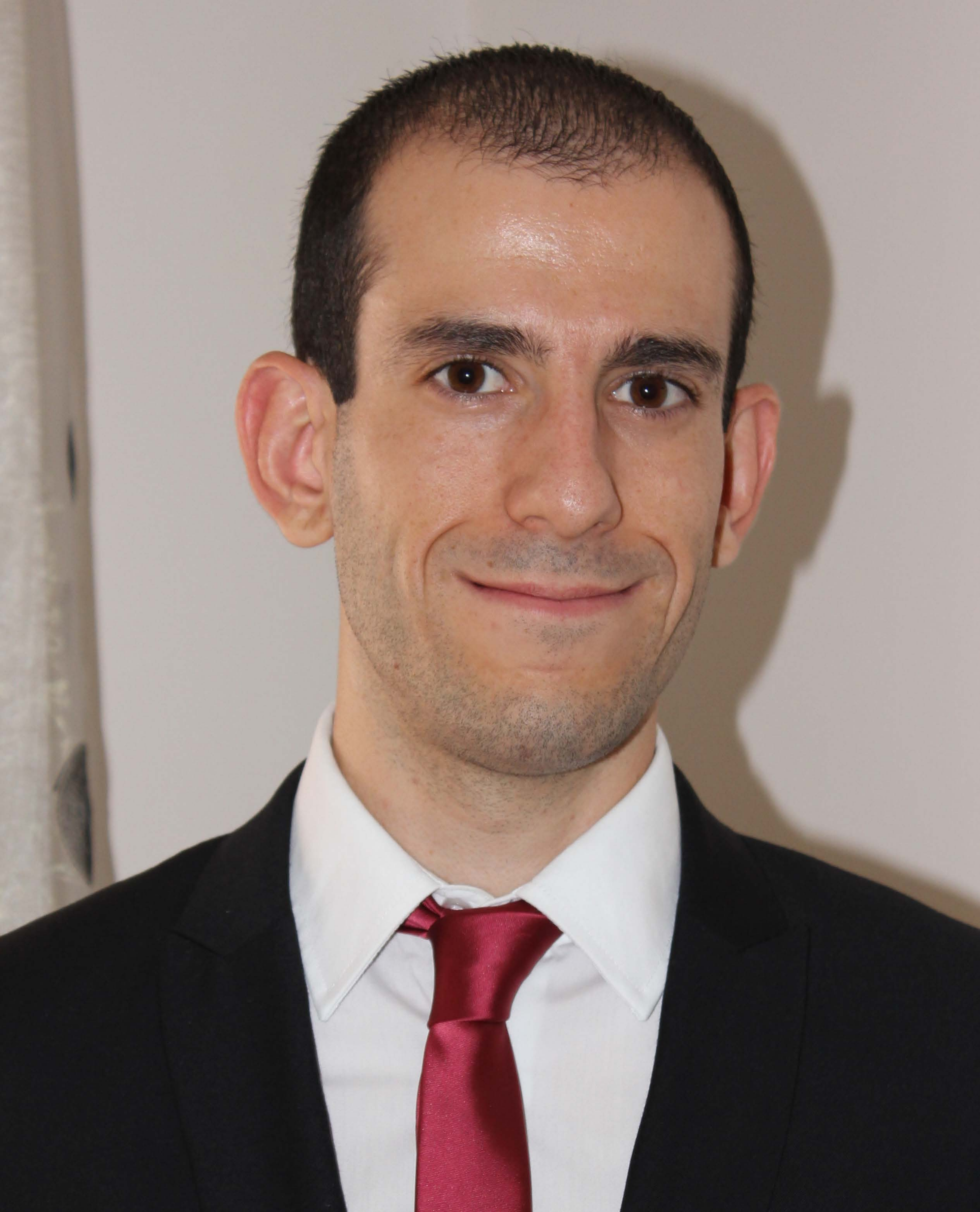}}]%
{Ruben Tolosana}
received the M.Sc. degree in Telecommunication Engineering, and his Ph.D. degree in Computer and Telecommunication Engineering, from Universidad Autonoma de Madrid, in 2014 and 2019, respectively. In 2014, he joined the Biometrics and Data Pattern Analytics - BiDA Lab at the Universidad Autonoma de Madrid, where he is currently collaborating as an Assistant Professor. Since then, Ruben has been granted with several awards such as the FPU research fellowship from Spanish MECD (2015), and the European Biometrics Industry Award (2018). His research interests are mainly focused on signal and image processing, pattern recognition, and machine learning, particularly in the areas of DeepFakes, HCI, and Biometrics. He is author of several publications and also collaborates as a reviewer in high-impact conferences (WACV, ICPR, ICDAR, IJCB, etc.) and journals (IEEE TPAMI, TCYB, TIFS, TIP, ACM CSUR, etc.). Finally, he is also actively involved in several National and European projects.
\end{IEEEbiography}

\begin{IEEEbiography}[{\includegraphics[width=1in,height=1.25in,clip,keepaspectratio]{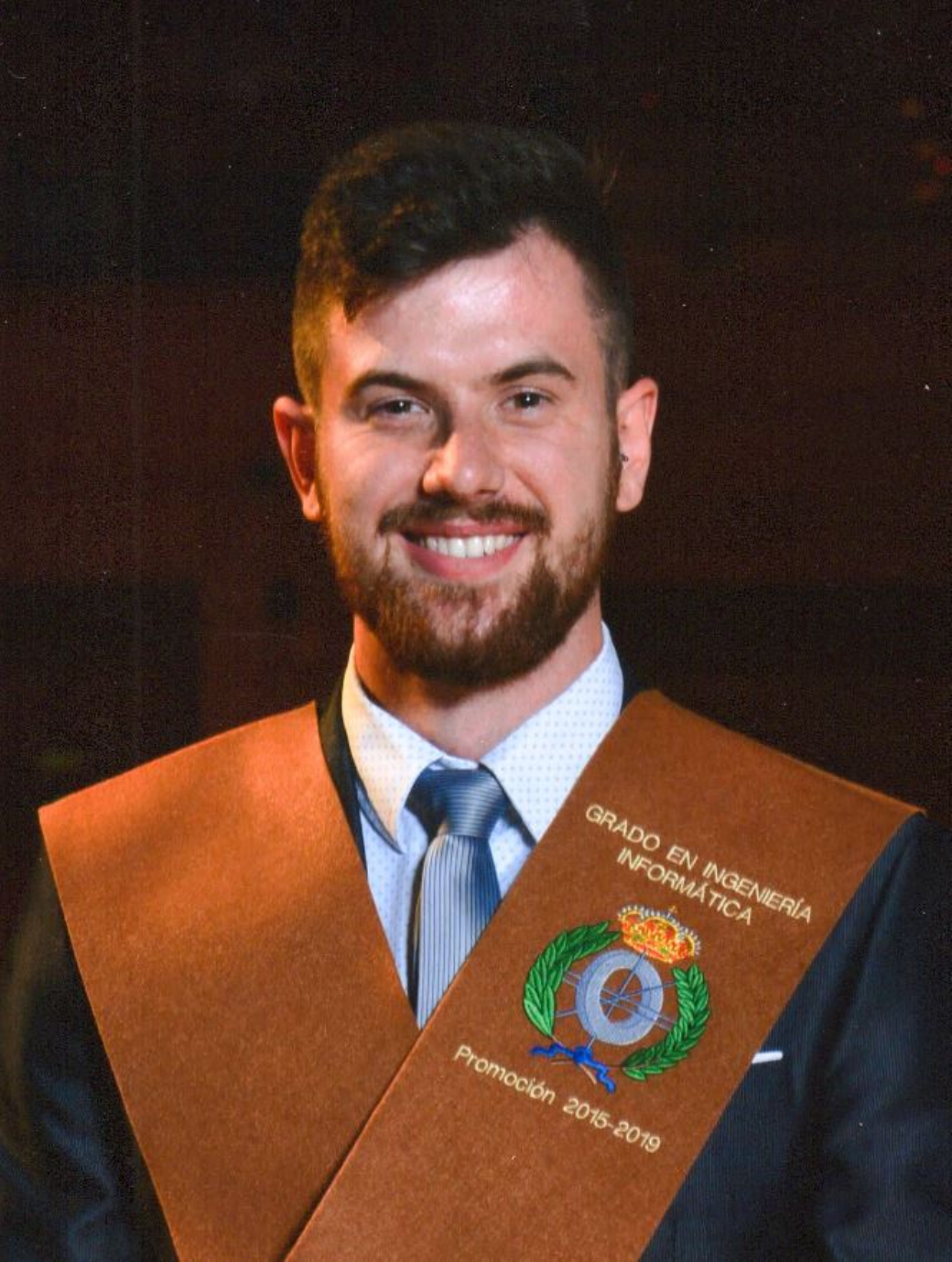}}]{Juan Carlos Ruiz-Garcia} received his B.Sc. degree in Computer Science Engineering from the Universidad de Granada, Spain, in 2019. He is currently studying the M.Sc. degree at the Universidad Autonoma de Madrid. In addition, in April 2020, he joined the Biometrics and Data Pattern Analytics - BiDA Lab as research assistant at the same university. His research focuses on the use of machine learning for e-Learning and e-Health.
\end{IEEEbiography}

\begin{IEEEbiography}[{\includegraphics[width=1in,height=1.25in,clip,keepaspectratio]{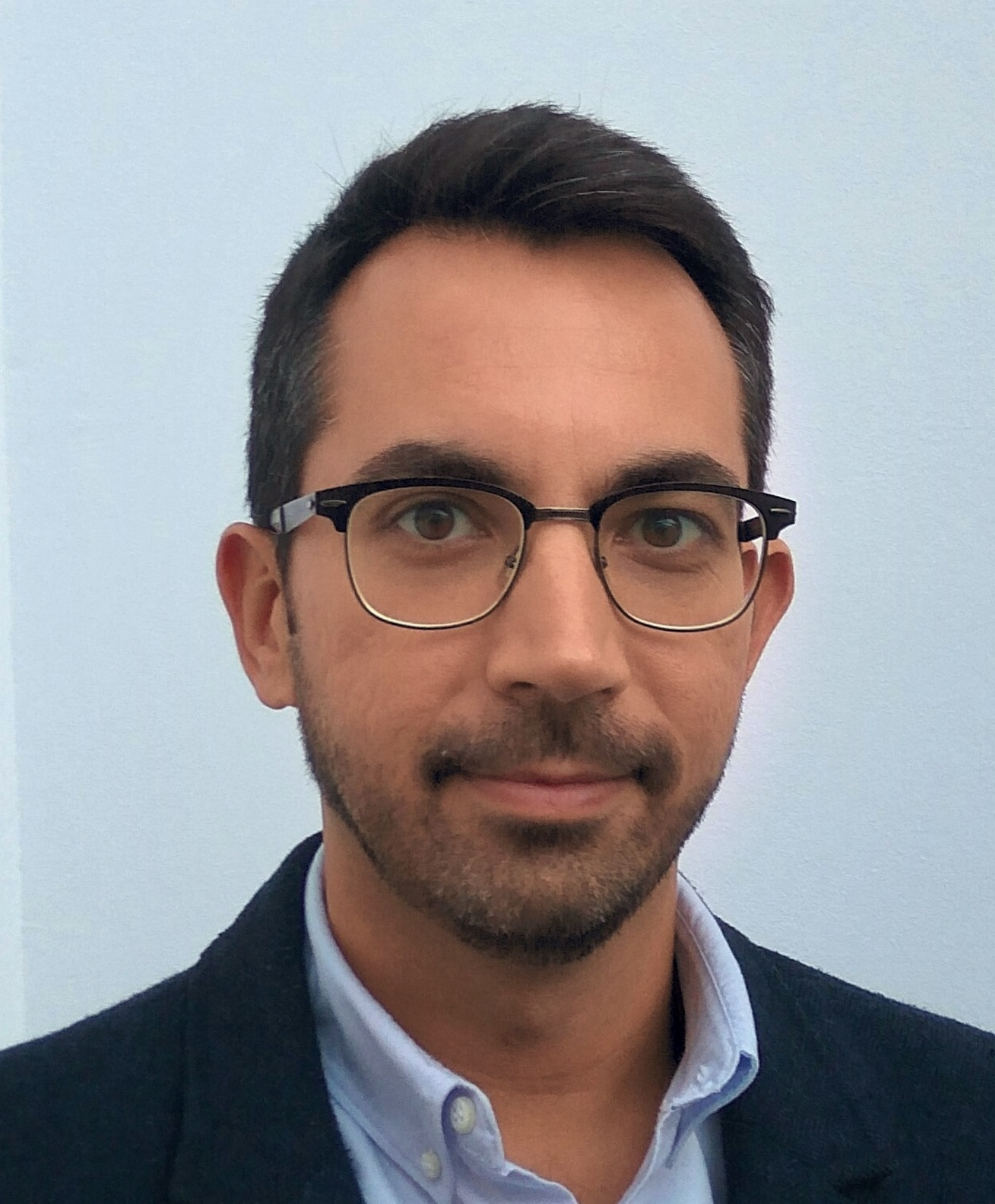}}]{Ruben Vera-Rodriguez} received the M.Sc. degree in telecommunications engineering from Universidad de Sevilla, Spain, in 2006, and the Ph.D. degree in electrical and electronic engineering from Swansea University, U.K., in 2010. Since 2010, he has been affiliated with the Biometric Recognition Group, Universidad Autonoma de Madrid, Spain, where he is currently an Associate Professor since 2018. His research interests include signal and image processing, pattern recognition, HCI, and biometrics, with emphasis on signature, face, gait verification and forensic applications of biometrics. Ruben has published over 100 scientific articles published in international journals and conferences. He is actively involved in several National and European projects focused on biometrics. Ruben has been Program Chair for the IEEE 51st International Carnahan Conference on Security and Technology (ICCST) in 2017; the 23rd Iberoamerican Congress on Pattern Recognition (CIARP 2018) in 2018; and the International Conference on Biometric Engineering and Applications (ICBEA 2019) in 2019.
\end{IEEEbiography}

\begin{IEEEbiography}[{\includegraphics[width=1in,height=1.25in,clip,keepaspectratio]{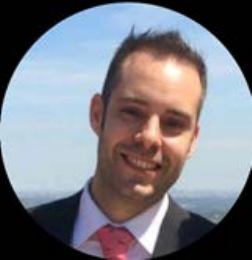}}]{Jaime Herreros-Rodriguez (JHR)} received the degree in Medicine in 2006 from Universidad Autónoma de Madrid, the tittle of neurologist in 2010 and he was awarded the title of Doctor in Medicine from the Universidad Complutense de Madrid (2019) with a distinction Cum Laude given unanimously for his doctoral thesis on migraine. He is also author of several publications in migraine and parkinsonism. He has collaborated with different research projects related to many neurological disorders, mainly Alzheimer and Parkinson's disease. JHR is a neurology and neurosurgery proffesor in CTO group, since 2008.
\end{IEEEbiography}

\begin{IEEEbiography}[{\includegraphics[width=1in,height=1.25in,clip,keepaspectratio]{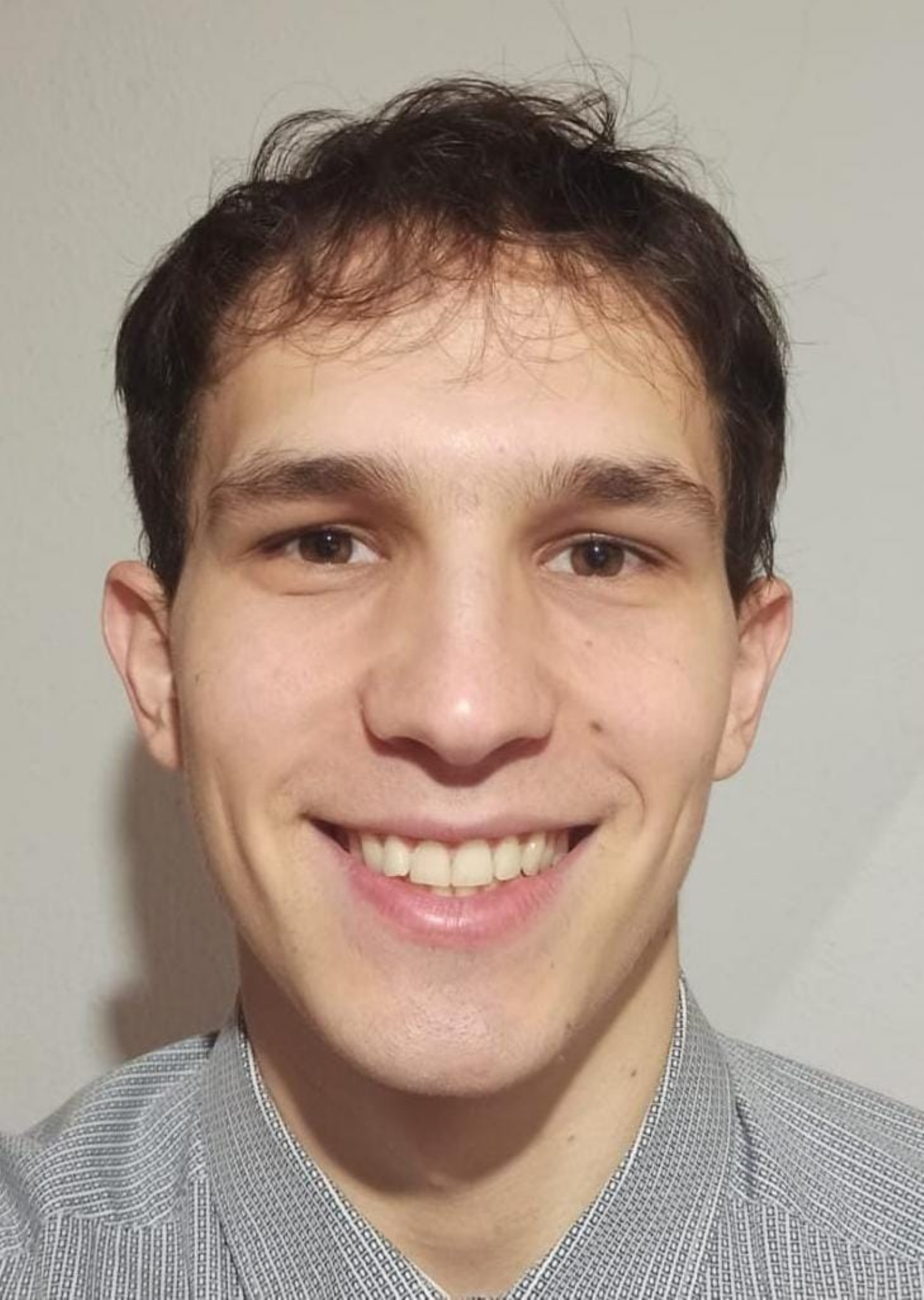}}]{Sergio Romero-Tapiador} received the BSc in Computer Science and Engineering in 2020 from Universidad Autonoma de Madrid. Since September 2019, he is a member of the Biometrics and Data Pattern Analytics - BiDA Lab at the Universidad Autonoma de Madrid, where he is currently collaborating as an assistant researcher. Among the research activities, he is mainly working on Pattern Recognition and Machine Learning, particularly in the area of DeepFakes.
\end{IEEEbiography}

\begin{IEEEbiography}[{\includegraphics[width=1in,height=1.25in,clip,keepaspectratio]{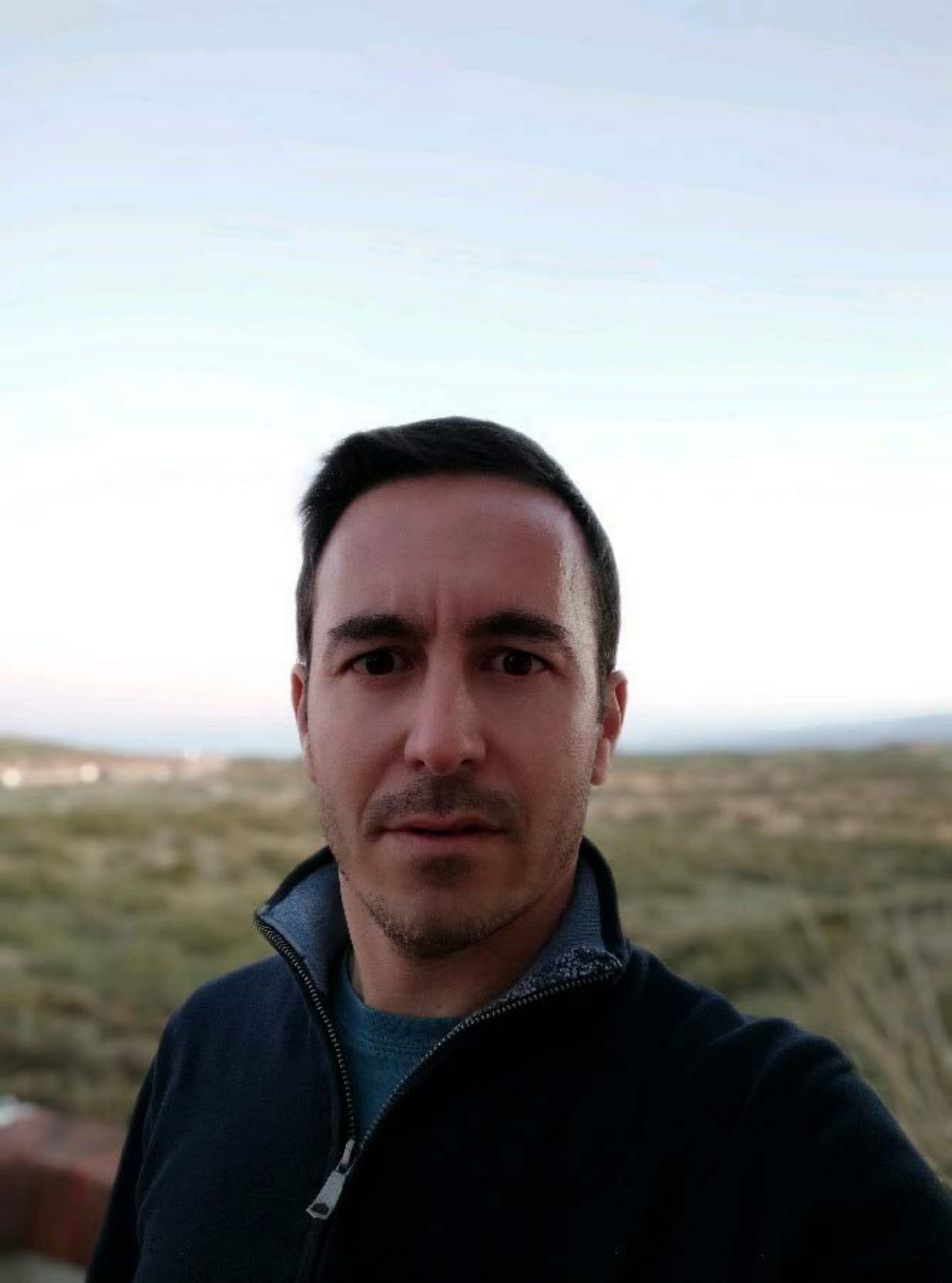}}]{Aythami Morales} received his M.Sc. degree in Telecommunication Engineering in 2006 from ULPGC. He received his Ph.D degree from ULPGC in 2011. He performs his research works in the BiDA Lab at Universidad Autónoma de Madrid, where he is currently an Associate Professor. He has performed research stays at the Biometric Research Laboratory at Michigan State University, the Biometric Research Center at Hong Kong Polytechnic University, the Biometric System Laboratory at University of Bologna and Schepens Eye Research Institute. His research interests include pattern recognition, machine learning, trustworthy AI, and biometrics. He is author of more than 100 scientific articles published in international journals and conferences, and 4 patents. He has received awards from ULPGC, La Caja de Canarias, SPEGC, and COIT. He has participated in several National and European projects in collaboration with other universities and private entities such as ULPGC, UPM, EUPMt, Accenture, Unión Fenosa, Soluziona, BBVA… 
\end{IEEEbiography}

\begin{IEEEbiography}[{\includegraphics[width=1in,height=1.25in,clip,keepaspectratio]{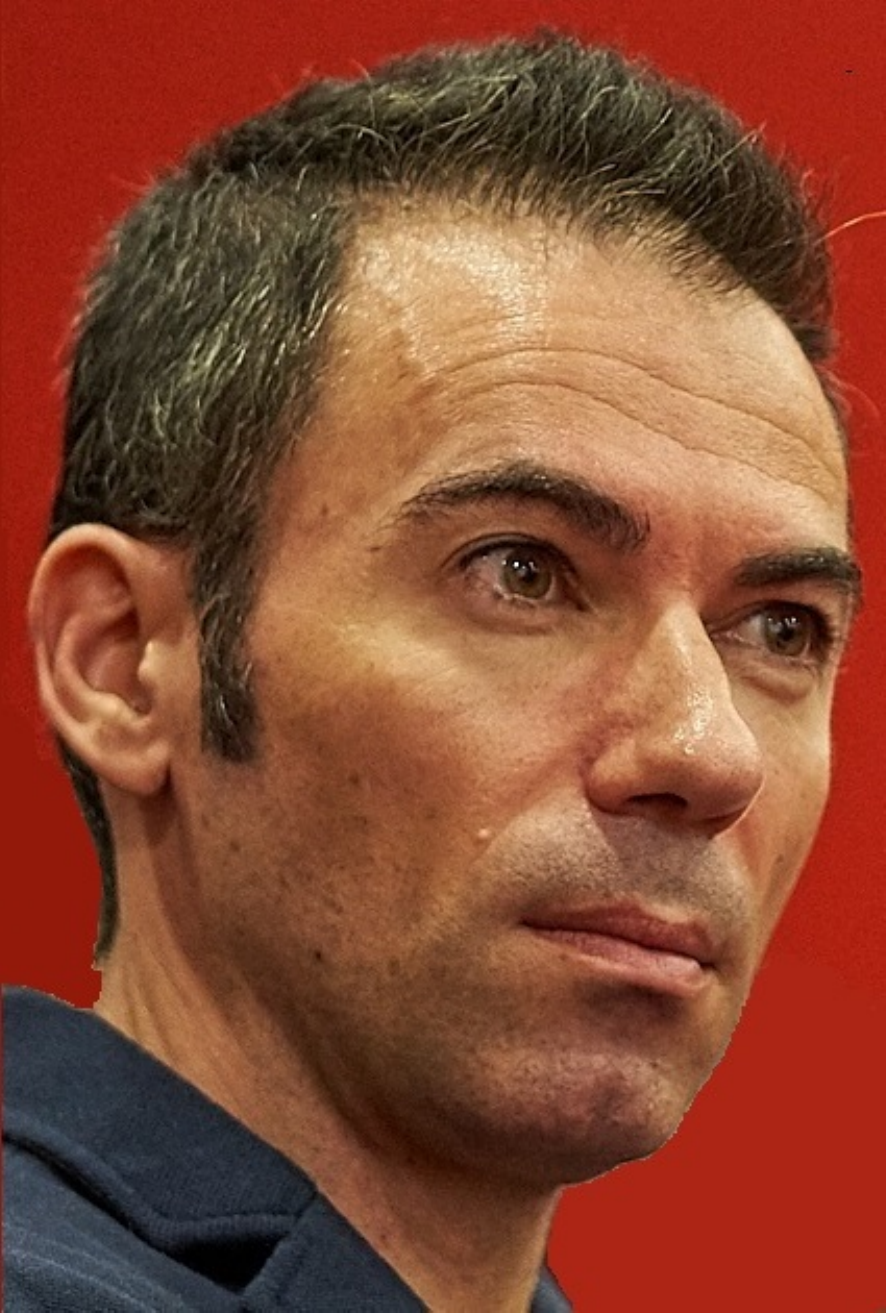}}]{Julian Fierrez} received the MSc and the PhD degrees from Universidad Politecnica de Madrid, Spain, in 2001 and 2006, respectively. Since 2004 he is at Universidad Autonoma de Madrid, where he is Associate Professor since 2010. His research is on signal and image processing, AI fundamentals and applications, HCI, forensics, and biometrics for security and human behavior analysis. He is Associate Editor for Information Fusion, IEEE Trans. on Information Forensics and Security, and IEEE Trans. on Image Processing. He has received best papers awards at AVBPA, ICB, IJCB, ICPR, ICPRS, and Pattern Recognition Letters; and several research distinctions, including: EBF European Biometric Industry Award 2006, EURASIP Best PhD Award 2012, Miguel Catalan Award to the Best Researcher under 40 in the Community of Madrid in the general area of Science and Technology, and the IAPR Young Biometrics Investigator Award 2017. Since 2020 he is member of the ELLIS Society.
\end{IEEEbiography}

%
%
%\begin{IEEEbiography}[{\includegraphics[width=1in,height=1.25in,clip,keepaspectratio]{Ortega.pdf}}]%
%{Javier Ortega-Garcia}
%received the M.Sc. degree in electrical engineering and the Ph.D. degree (cum laude) in electrical engineering from Universidad Politecnica de Madrid, Spain, in 1989 and 1996, respectively. He is currently a Full Professor at the Signal Processing Chair in Universidad Autonoma de Madrid - Spain, where he holds courses on biometric recognition and digital signal processing. He is a founder and Director of the BiDA-Lab, Biometrics and Data Pattern Analytics Group. He has authored over 300 international contributions, including book chapters, refereed journal, and conference papers. His research interests are focused on biometric pattern recognition (on-line signature verification, speaker recognition, human-device interaction) for security, e-health and user profiling applications. He chaired Odyssey-04, The Speaker Recognition Workshop, ICB-2013, the 6th IAPR International Conference on Biometrics, and ICCST2017, the 51st IEEE International Carnahan Conference on Security Technology.
%\end{IEEEbiography}

\end{document}